\documentclass[11pt,a4paper]{article}

\usepackage {amsmath, amssymb, amsfonts,bm}
\usepackage {graphics}
\usepackage {graphicx}
\usepackage{natbib,url}

\usepackage {longtable}
\usepackage {float,afterpage,color} % for putting figure exactly where you want to, see appendix here
\usepackage {verbatim} % for begin comment, end comment
\usepackage {subfigure}
\usepackage{chemarr} % for xrightleftharpoons
\usepackage{lscape}
\usepackage{bbm,dsfont}
\usepackage{setspace}
\usepackage{fancyheadings}
\usepackage{float,afterpage}
\usepackage{multirow}
\usepackage{amsthm}
\usepackage{color}
\usepackage {graphics}
\usepackage {graphicx}
\usepackage{pgf,tikz,float,pgfpages}
\usepackage{amsmath,amssymb,hyphenat,bbold,amstext}
\usetikzlibrary{arrows,snakes,backgrounds,matrix,topaths,petri,fit,trees,patterns,shapes,shadows,topaths}
\usetikzlibrary{plothandlers,plotmarks}
\usetikzlibrary{decorations.pathmorphing,positioning}

\usepackage{anysize}
\marginsize{2cm}{2cm}{2cm}{2cm}

\DeclareMathOperator{\argmax}{argmax}

\newcommand{\1}{\mathds{1}}

\newtheorem{remark}{Remark}

\newcommand{\eg}{e.g.\ }

\newcommand{\be}{\begin{equation}}
\newcommand{\ee}{\end{equation}}

\newcommand{\Qterm}{Q(K_t,\epsilon_{t-1}, \epsilon_t,x)}
\newcommand{\alphaterm}{\alpha(K_t,\epsilon_{t-1},\epsilon_t,x)}
\newcommand{\Cterm}{C(\epsilon_{t-1},\epsilon_t,x)}

\usepackage{algorithm}
\usepackage{algorithmicx,algpseudocode}

\title{On optimality of kernels for approximate Bayesian computation using sequential Monte Carlo}
\author{Sarah Filippi$^1$, Chris P. Barnes$^1$, Julien Cornebise, Michael P.H. Stumpf$^1$ \\ \small $^1$Centre for Integrative Systems Biology and Bioinformatics,\\ \small Imperial College London, London SW7}

\begin{document}

\maketitle
\begin{abstract}
Approximate Bayesian computation (ABC) has gained popularity over the past few years for the analysis of complex models arising in population genetic, epidemiology and system biology. 
Sequential Monte Carlo (SMC) approaches have become work horses in ABC. Here we discuss how to construct the perturbation kernels that are required in ABC SMC approaches, in order to construct a set of distributions that start out from a suitably defined prior and converge towards the unknown posterior. We derive optimality criteria for different kernels, which are based on the Kullback-Leibler divergence between a distribution and the distribution of the perturbed particles. We will show that for many complicated posterior distributions, locally adapted kernels tend to show the best performance. In cases where it is possible to estimate the Fisher information we can construct particularly efficient perturbation kernels. We find that the added moderate cost of adapting kernel functions is easily regained in terms of the higher acceptance rate. We demonstrate the computational efficiency gains in a range of toy-examples which illustrate some of the challenges faced in real-world applications of ABC, before turning to two demanding parameter inference problem in molecular biology, which highlight the huge increases in efficiency that can be gained from choice of optimal models. We conclude with a general discussion of rational choice of perturbation kernels in ABC SMC settings.
\end{abstract}

\section{Introduction}

Statistical practice and theory tend to reflect scientific fashions \citep{Stigler:1986aa}. Today mathematical models in biological sciences are becoming increasingly complex. This, together with the deluge of data being produced typically in genetics and genomics, poses severe challenges to statistical inference \citep{Efron:2010aa}. In particular in many area of computational biology evaluation of the likelihood \citep{Cox2006}
$$
L(\theta) = f(x|\theta),
$$
where $x$ are realizations of the data, and $\theta$ is the (potentially vector-valued) parameter characterizing the data-generating process, is often turning out to be impractical. 
Approximate Bayesian Computation (ABC) methods \citep{Beaumont:2002ue,Marin:2011ug} were first conceived to allow (Bayesian) statistical inference in situations where the evaluation of the likelihood is too complicated or numerically too demanding \citep{Pritchard:1999td,Tanaka:2006ga,Lopes:2010ea}. Rather than evaluating the likelihood directly, ABC-based approaches use systematic comparisons between real and simulated data in order to arrive at approximations of the true (but unobtainable) posterior distribution,
$$
p(\theta|x) \propto  f(x|\theta)\pi(\theta),
$$
where $\pi(\theta)$ denotes the prior distribution of $\theta$. 
\par
Simulating from $f(x|\theta)$ is generally straightforward, even if obtaining a reliable numerical/functional representation of the model is not possible. We then compare the simulated data, $y$, with the real data, $x$, and accept only those simulations where some distance measure between the two, $\Delta(x,y)$,  falls below a specified threshold, $\epsilon$. If the data are too intricate or complicated it is common to replace a comparison of the real and simulated data by a comparison of suitable summary statistics. This results in a often appreciable reduction of the dimension but is fraught with problems if the summary statistics are not sufficient. Given that sufficiency \citep{Cox2006} is a rare quality indeed \citep{Lehmann:1993aa}, and probably not given for any real-world problem of scientific interest, this problem is now attracting a lot of attention \citep{Robert:2011wz,Didelot:2010wo,Fearnhead:2010vj}. Here, however, we shall focus on the data directly; we thus seek to determine approximate posteriors of the form,
$$
p(\theta|x)\approx p_\epsilon(\theta|x)\propto\int f(y|\theta)\; \mathds{1}\left(\Delta(x,y)\le \epsilon\right) \pi(\theta) dy,
$$
where $y$ is the data simulated from the model $f(\cdot|\theta)$ for a given parameter, $\theta$, drawn from the appropriate prior distribution, and $x$ is the observed data.
\par
The simple ABC scheme outlined above suffers from the same shortcomings as other rejection samplers: most of the samples are drawn from regions of parameter space, which cannot give rise to simulation outputs that resemble the data. Therefore a number of computational schemes have been proposed that makes ABC inference more efficient. These come in loosely three flavours: regression-adjusted ABC \citep{Tallmon:2004jp,Fagundes:2007fr,Blum:2010hv}, Markov chain Monte Carlo ABC schemes \citep{Marjoram:2003fn,Ratmann:2007hh}, and ABC implementing some variant of sequential importance sampling (SIS) or sequential Monte Carlo (SMC) \citep{Sisson:2007aa,Toni:2009p9197,Beaumont:2009be,DelMoral:2008tm}. Of these flavours the first and the last forms have received the greatest attention and it is an ABC scheme based on sequential importance sampling that we will focus on as it offers greater flexibility and applicability and appears to be enjoying greater popularity in applications.
\par
We focus on the implementation of \cite{Toni:2009p9197} and \cite{Beaumont:2009be} (called ABC SMC in the following), which like other related SIS and SMC methods works by constructing a series of intermediate distributions that start out from a suitably specified prior distribution and increasingly resemble the (unknown) approximate posterior distribution. These methods aim to sample sequentially from a sequence of distributions, which increasingly resemble the target posterior; they are constructed by estimating intermediate distributions $p_{\epsilon_t}(\theta|x)$ for a decreasing sequence of $\{\epsilon_t\}_{1\leq t\leq T}$. Each intermediate distribution is described by a weighted sample of parameter vectors. Successive distributions are constructed by sampling parameters from the previous population, perturbing them through some kernel function, $\tilde{\theta}\sim K_t(\cdot|\theta)$, generating simulated data, $y\sim f(\cdot|\tilde{\theta})$ and, upon acceptance, calculating the corresponding new weights. 
\par
While this sequential ABC approach is computationally much more efficient than simple ABC rejection schemes, the overall computational burden does not only depend on the complexity of the model and the amount of data at hand, but also on details of the chosen sequential scheme. In particular the $\epsilon$-schedule, $\{\epsilon_1,\ldots,\epsilon_T\}$, and the choice of perturbation kernels, $K_t(\cdot|\cdot)$ exert considerable influence on the algorithmic complexity. As in many Monte Carlo settings \citep{Gilks1996,Robert2004} problems tend to arise as the dimension of the parameter space increases and balancing convergence with an exhaustive exploration of the parameter space becomes harder. In this paper we focus on perturbation kernel selection and its effect on the algorithm efficiency.
\par
The construction of suitable kernel functions has been a longstanding problem in importance sampling \citep{ohBerger1993,Givens:1996tc}, sequential importance sampling and population Monte-Carlo \citep{Douc:2007ue,Cappe:2008ht,Cornuet:2009vz} as well as in sequential Monte-Carlo for state space models \citep{Pitt1999b,Der_Merwe2001,Cornebise:2008}.
However, it is still far from being solved especially in an ABC context where formal and informal understanding of kernel choice remain areas of pressing concern.
\par
Especially for models which are computationally expensive to simulate, such as dynamical systems \citep{Gutenkunst:2007,Secrier:2009ko,Erguler:2011bu}, the choice of the kernel will have huge influence on the efficiency with which parameter spaces are explored and posterior estimates obtained. Here we will discuss a range of kernel functions, characterize their performance, and put forward some analytic results as to their optimality. In the next section we discuss the ABC scheme in some detail before describing criteria for optimally choosing the perturbation kernels and outlining different classes of perturbation kernels. We  then examine the performance of these kernels in a range of illustrative problems and compare their algorithmic complexity. We will then show that for two models in molecular biology with complex posterior parameter distributions the choice of suitable kernels can vastly improve the computational cost of ABC SMC inferences.

\section{The ABC SMC algorithm}
The general scheme of ABC inference is as follows:
\begin{itemize}
\item sample a parameter vector $\theta$ (also called a \textit{particle}) from the prior distribution $\pi(\theta)$,
\item simulate a dataset $y$ according to the generative model $f(y|\theta)$,
\item compare the simulated dataset with the experimental data $x$: if $\Delta(x,y)\le\epsilon$, accept the particle.
\end{itemize}
This scheme is repeated until $N$ particles are accepted; these form a sample from the posterior distribution $$p_\epsilon(\theta|x) \propto \int \mathds{1}(\Delta(x,y)\le \epsilon) f(y|\theta)\pi(\theta)dy$$ which is an approximation of the posterior distribution $p(\theta|x)$.
\par
Over the past few years many improvements of these algorithms have been proposed. In particular, \cite{Marjoram:2003fn} introduced a method based on Markov chain Monte Carlo, which consists in constructing a Markov chain whose stationary distribution is $p_\epsilon(\theta|x)$. 
%To do so, at each time $t$, a particle $\theta$ is simulated from the previous particle $\theta^{(t-1)}$ according to a perturbation kernel $K(\cdot|\theta^{(t-1)})$ (or from the prior distribution for $t=0$); the simulated data $y\sim f(\cdot|\theta)$ is compared with the experimental data; and $\theta^{(t)}$ is set to be equal to $\theta$ with a Metropolis Hasting acceptance rate if $\Delta(y,x)\le \epsilon$ and to $\theta^{(t-1)}$ otherwise. 
This algorithm is guaranteed to converge, however, it is very difficult to assess when the Markov chain reaches the stationary regime; furthermore the chain may get trapped in local extrema.
\par 
 SIS and SMC samplers have then been introduced in the ABC framework by several authors \citep{Sisson:2007aa,Toni:2009p9197,Beaumont:2009be,DelMoral:2008tm}. These methods aim to sample sequentially from the distributions $p_{\epsilon_t}(\theta|x)$ for a decreasing sequence of $\{\epsilon_t\}_{1\leq t\leq T}$. The scheme of the algorithm is as follows: first, the ABC algorithm described above is used to construct a sample from $p_{\epsilon_1}(\theta|x)$ with a sufficiently large value of $\epsilon_1$ such that many particles are accepted. The ABC algorithm is then used again with $\epsilon_2$ as a threshold; but instead of sampling parameters from the prior, they are sampled from the set of accepted particles at the previous stage and perturbed according to a suitable \textit{perturbation kernel}. This way a sample from $p_{\epsilon_2}(\theta|x)$ is built, and so on until our target posterior has been reached.
 In this article, we focus on the implementation of \cite{Toni:2009p9197} and \cite{Beaumont:2009be} described in Algorithm~\ref{algo:ABCSMC}. In the following we will refer to this implementation as the ABC SMC algorithm according to \cite{Toni:2009p9197}. The ABC Population Monte Carlo algorithm proposed by \cite{Beaumont:2009be} is similar to the ABC SMC algorithm, except that a specific perturbation kernel is used.  It is, however, worth distinguishing between these algorithms and the one of \cite{DelMoral:2008tm} and \cite{Drovandi:2011hm} based on the SMC sampler of \cite{DelMoral:2006wv}. When using SMC samplers, both a forward and a backward kernel need to be defined, which reduces the algorithmic complexity from $O(N^2)$ to $O(N)$ where $N$ is the number of particles. However, in many applications of interest, the most computationally expensive part of an ABC algorithm is the simulation of the data, which although of complexity $O(N)$ dominates the $O(N^2)$ term for any practically feasible value of $N$ \citep{Beaumont:2009be}. In this paper, we will not discuss kernel choice in the context of these approaches, although here, too, the choice of kernel will impact  the numerical efficiency. 
\begin{algorithm}[h!]
 \begin{algorithmic}[1]
   \caption{ABC SMC algorithm}
   \label{algo:ABCSMC}
   \State {\bfseries input:} a threshold $\epsilon$
   \State {\bfseries output:} a weighted sample of particles from $p_{\epsilon}(\theta|x)$
   \State $t \leftarrow 0$
   \Repeat
   \State  $t \leftarrow t+1$
   \State determine the next threshold $\epsilon_t$
   \State determine the parameters of the perturbation kernel $K_t(\cdot|\cdot)$
   \State $i \leftarrow 1$
   \Repeat
   \If{t=1}
   \State sample $\tilde\theta$ from $\pi(\theta)$
   \Else
   \State sample $\theta$ from the previous population $\{\theta^{(i,t-1)}\}_{1\leq i\leq N}$ with weights $\{\omega^{(i,t-1)}\}_{1\leq i\leq N}$
   \State sample $\tilde\theta$ from $K_t(\cdot|\theta)$ and such that $\pi(\tilde\theta)>0$
   \EndIf
   \State sample $y$ from $f(\cdot|\tilde\theta)$
   \If {$\Delta(y,x)\le\epsilon_t$}
   \State $\theta^{(i,t)}\leftarrow\tilde\theta$
   \State $y^{(i,t)}\leftarrow y$
   \State $i\leftarrow i+1$
   \EndIf
   \Until{ $i=N+1$}
   \State calculate the weights: for all $1\leq i\leq N$
   \If{$t\neq 1$}
   $$\omega^{(i,t)} \leftarrow\frac{\pi(\theta^{(i,t)})}{\sum_{j=1}^n \omega^{(j,t-1)}K_t(\theta^{(i,t)}|\theta^{(j,t-1)})}$$
   \Else $\quad\omega^{(i,1)}\leftarrow1$
   \EndIf
   \State normalize the weights
  \Until{ $\epsilon_t\leq \epsilon$}
  \State $T\leftarrow t$
 \end{algorithmic}
\end{algorithm}
\par
The behaviour of the algorithm depends on its settings: in particular the decreasing sequence of $\{\epsilon_t\}_t$ and the perturbation kernels $\{K_t(\cdot|\cdot)\}_t$. The effect of the sequence of decreasing threshold is easy to understand: if the difference between two successive tolerances $\epsilon_t$ and $\epsilon_{t+1}$ is small, the posterior distributions $p_{\epsilon_t}(\theta|x)$ and $p_{\epsilon_{t+1}}(\theta|x)$ are similar and a small number of simulations will be required to generate $N$ draws from the next intermediate distribution, $p_{\epsilon_{t+1}}(\theta|x)$, by sampling from the weighted population $\{\theta^{(i,t-1)},\omega^{(i,t-1)}\}_{1\leq i\leq N}$. But a slowly decreasing sequence of thresholds $\{\epsilon_t\}_{1\leq t\leq T}$ leads to a large number of iterations (large value of $T$) in order to obtain $\epsilon_T=\epsilon$. In practise, until recently, the sequence of tolerance thresholds were most often tuned by hand according to the model. An adaptive choice of the threshold schedule has been proposed by \cite{DelMoral:2008tm} and \cite{Drovandi:2011hm}. It consists in selecting the $\alpha$-th quantile of the distances between the simulated data $\{y^{(i,t)}\}_{1\leq i \leq N}$ and the observed one $x$. Selecting the threshold adaptively often significantly improves the efficiency of the ABC SMC algorithm. However, the efficiency strongly depends on the choice of $\alpha$ and as it is argued in \cite{Silk:Filippi:2012} for some values of $\alpha$ the algorithm may not converge to the posterior distribution $p_\epsilon(\cdot|x)$.
\par
 Similarly, the choice of the perturbation kernels $\{K_t(\cdot|\cdot)\}_{1\leq t\leq T}$ exerts considerable influence on the computational complexity of the algorithm. A local perturbation kernel hardly moves the particles and has the advantage to produce new particles which are accepted with high probability if the successive values of $\epsilon$ are close enough; on the other hand, a widely spread out or permissive perturbation kernel enables exploring the parameter space more fully, but does so at the cost of achieving only low acceptance rates.

\section{Properties of optimal kernels}
\newcommand{\proposal}{q_{\epsilon_{t-1},\epsilon_t}}
\newcommand{\target}{q^*_{\epsilon_{t-1},\epsilon_t}}
\newcommand{\KLD}{KL}
\newcommand{\old}{\theta^{(t-1)}}
\newcommand{\new}{\theta^{(t)}}

In sequential importance sampling, a perturbation kernel $K_t$ should fulfill several requirements to be computationally efficient. In particular, the joint proposal distribution, corresponding to picking a particle at
 random and perturbing it to obtain a new particle, should ``resemble'' in some sense the
 target joint distribution, corresponding to  picking independently
 two particles. 
 More precisely, the joint proposal distribution of a particle, which samples
 first a particle $\theta^{(t-1)}\sim p_{\epsilon_{t-1}}(\cdot|x)$ then a
 perturbed particle $\theta^{(t)}\sim K_t(\cdot| \theta^{(t-1)})$, and accept the couple if and
 only if $\Delta(y,x)\leq\epsilon_t$ where $y\sim f(\cdot|\theta^{(t)})$,  admits for
 density
 \begin{equation}
q_{\epsilon_{t-1},\epsilon_t}(\theta^{(t-1)},\theta^{(t)}|x)=
\frac{p_{\epsilon_{t-1}}(\theta^{(t-1)}|x)K_t(\theta^{(t)}|\theta^{(t-1)})\int f(y|\theta^{(t)})\; \mathds{1}\left(\Delta(x,y)\le \epsilon_t\right)dy}
{\alpha(K_t,\epsilon_{t-1},\epsilon_t,x)}
\;.
\label{eq:jointproposal}
\end{equation}
The normalization factor 
\begin{equation}
\alpha(K_t,\epsilon_{t-1},\epsilon_t,x)
= \iiint p_{\epsilon_{t-1}}(\theta^{(t-1)}|x)K_t(\theta^{(t)}|\theta^{(t-1)})
f(y|\theta^{(t)})\; \mathds{1}\left(\Delta(x,y)\le \epsilon_t\right) d\old
d\new dy
\label{eq:defalpha}
\end{equation}
 is the \emph{average acceptance probability}, that is, the
 proportion of proposed particles that are not rejected. 
 This joint proposal distribution should ``resemble'' in some sense the target
 product distribution, that of sampling $\theta^{(t-1)}$ and $\theta^{(t)}$ independently from, respectively, $p_{\epsilon_{t-1}}(\cdot|x)$ and
$p_{\epsilon_{t}}(\cdot|x)$, whose density is
\begin{equation}
q^*_{\epsilon_{t-1},\epsilon_t}(\theta^{(t-1)},\theta^{(t)}|x)=p_{\epsilon_{t-1}}(\theta^{(t-1)}|x)p_{\epsilon_t}(\theta^{(t)}|x)\;.
\label{eq:jointtarget}
\end{equation}
As argued by several authors, \eg
~\cite{Douc:2007ue,Cappe:2008ht,Cornebise:2008,Beaumont:2009be}, a
mathematically convenient formal definition of this ``resemblance'' is the
Kullback-Leibler (KL) divergence between the proposal distribution
$\proposal(\old, \new)$ and the target distribution $\target(\old, \new)$, i.e.
\begin{equation}
	\KLD(\proposal;\target) 
	 = \iint \target\left(\old, \new\right) \log\frac{\target\left( \old, \new \right)}{\proposal\left(\old,\new \right)} d \old d \new \;.
	 \label{eq:KLDjoint}
\end{equation}
In order to determine the variance of a component-wise Gaussian kernel, \cite{Beaumont:2009be} also minimized this quantity, albeit considering
only the special case where $\epsilon_{t-1} = \epsilon_t = 0$ for which the solution has a closed form. In particular, in this special case, $\alpha(K_t,0,0,x) = 1$ for any $K_t$.
\par
However, in addition to maximize $\KLD(\proposal;\target)$, an efficient ABC SMC proposal kernel $K_t$ should have a high average acceptance rate $\alphaterm$. When refining \cite{Beaumont:2009be} with $\epsilon_t\neq \epsilon_{t-1}$, we remark that $\KLD(\proposal;\target)$ can be separated into three terms

\begin{equation}
	\KLD(\proposal;\target) 
	 = - Q(K_t, \epsilon_{t-1}, \epsilon_t,x) + \log \alphaterm +
	 \Cterm \label{eq:KLDterms}\;,
\end{equation}
where
\begin{equation}
\Qterm = \iint 
	p_{\epsilon_{t-1}}(\theta^{(t-1)}|x)p_{\epsilon_t}(\theta^{(t)}|x) \log
	K_t(\theta^{(t)}|\theta^{(t-1)}) d\theta^{(t-1)}d\theta^{(t)}	\label{eq:Qterm}\;,
\end{equation}
can  be maximized easily in some convenient cases (see
Section~\ref{sec:somekernels}), $\alphaterm$ is the average acceptance probability
already defined in \eqref{eq:defalpha}, which is much harder to minimize, 
and $\Cterm$ does not depend on the kernel $K_t$. 
Therefore  the two following maximization problems are equivalent:
\begin{equation}
	\argmax_{K_t}  \Qterm 
	= \argmax_{K_t} \left(- \KLD(\proposal;\target) + \log	\alphaterm \right)	\label{eq:smartMaximization} \;.
\end{equation}
As we will show in Section~\ref{sec:somekernels} this problem is easy to solve since the left-hand side often admits a
closed-form solution. The most important remark is that the
right-hand side is the solution of a \emph{multi-objective optimization
problem}, solving a trade-off between jointly \emph{minimizing the Kullback-Leibler divergence} and \emph{maximizing the logarithm of the average acceptance
probability}.  Multi-objective optimization using an additive
combination of two distinct objective functions is common practice, see e.g. Section 4.7.5 of \cite{Boyd2004}. Although
the weights of such additive combination are here forced upon us, we note that
the use of the logarithm of the acceptance probability strongly penalizes 
very low probabilities, while making equally desirable moderate to large acceptance probabilities,
a reasonable preference from the
computational point of view.
\par
%\cite{Beaumont:2009be} follow a similar line of reasoning, albeit considering
%only the asymptotic case where $\epsilon_{t-1} = \epsilon_t = 0$. Therefore,
%they minimize $\KLD(q_{0,0};q^\ast_{0,0})$, and, since $\alpha(K_t,0,0,x) = 1$ for any $K_t$, %they would ideally want to find the kernel that
%maximizes $ Q(K_t,0,0,x)$. Because it is impossible to solve this problem in
%the asymptotic case, they revert back to the non-asymptotic case by setting
%$\epsilon_t = \epsilon_{t-1}$ for both thresholds, and therefore eventually
%solve
%$\argmax_{K_t} 
%Q(K_t,\epsilon_{t-1},\epsilon_{t-1},x)$.
%In their study they restrain themselves to the component-wise Gaussian kernel (see %Section~\ref{sec:componentwise}).
%\par
%However, a close study of Equation~\eqref{eq:KLDterms} shows that this approach
%does not actually minimize the real Kullback-Leibler divergence in the
%non-asymptotic case, as we would hope to. Eventually minimizing this
%KL divergence would require two modifications: evaluate $\Qterm$ with the two distinct
%thresholds $\epsilon_{t-1} \neq \epsilon_t$, as we easily do in Section~\ref{sec:somekernels} %-- it simply amounts to replacing a covariance by a cross-covariance in the solution -- and, %much more troublesome, take into account $\log \alphaterm$. This second point is by and %arge impractical; there is no closed-form (that is, easily computable) solution to maximize the %cceptance-rate. Therefore, our original goal of minimizing $\KLD(\proposal;\target)$ per se %would seem unattainable.
\par
The practical consequence of these theoretical arguments is to choose the proposal kernel
\begin{equation}
K_t = \argmax_{K_t} \Qterm\;.
\label{eq:actualArgmax}
\end{equation} 
We have shown that this choice corresponds to a trade-off between two desirable properties of the
kernel, namely the resemblance of the proposal distribution and the
target in the sense of the KL divergence, and a high acceptance rate. And
our criterion not only sheds new light on the justification of some existing,
proven criteria, but, additionally, refines them. In the next section we will describe how to carry out this maximisation in practice from a set of particles for random walk kernels.

\begin{remark}[Resampling and finite population size]
	For the sake of clarity, we have slightly simplified the equations above by omitting the resampling step of ABC-SMC. In reality it is not possible to sample exactly $\theta^{(t-1)}$ from $p_{\epsilon_{t-1}}(\cdot|x)$; rather, as described in Algorithm~\ref{algo:ABCSMC}, $\theta^{(t-1)}$ is sampled with replacement from the previous population of particles $\{\theta^{(i,t-1)}\}$, eventually sampling from the weighted empirical distribution of the previous population, noted $p^N_{\epsilon_{t-1}}(\cdot|x)$. Systematically replacing $p_{\epsilon_{t-1}}(\cdot|x)$ with $p^N_{\epsilon_{t-1}}(\cdot|x)$ throughout equations~\eqref{eq:jointproposal} to~\eqref{eq:actualArgmax} leads to the exact optimality criterion that we used, properly defined for a finite population size $N$.
\end{remark}

\begin{remark}[Theoretical requirements for convergence]
		Convergence of importance sampling-based algorithms such as SMC samplers and ABC-SMC places some precise conditions on the kernel: i) having a larger support than the target distribution, to guarantee asymptotic unbiasedness of the empirical mean by a law of large numbers; and ii) vanishing slowly enough in the tails of the target, equivalent to ensuring finite variance of the importance weights, to guarantee asymptotic normality and finite variance of the estimator by a central limit theorem. 

		Fully investigating whether the optimal kernels satisfy such conditions is outside the scope of this article. However, the Gaussian kernels studied  in the following sections have unbounded support, therefore invoking a law of large numbers. The direction of the asymmetric KLD that we consider in equation~\eqref{eq:KLDjoint} facilitates (but does not guarantee) finite variance. Indeed, this direction is opposite to that of the KLD used in Variational Bayes, but similar to that used in Expectation Propagation algorithms \citep{barthelme2011expectation}. As explained by e.g. \citep[p. 468]{bishop2006pattern}, this direction favours proposals with a good coverage of the target, therefore heuristically tending to satisfy the requirements for asymptotic normality.
		
\end{remark}

\section{Optimal choice of random walk kernels}\label{sec:somekernels}
Perturbing a particle $\theta^{(j,t-1)}$ consists of generating a new particle according to a probability parametrized by $\theta^{(j,t-1)}$ and often centered on $\theta^{(j,t-1)}$. In the ABC SMC algorithm, in addition to sampling from the kernel, we must be able to compute the transition density $K_t(\theta^{(i,t)}|\theta^{(j,t-1)})$ for any particles $\theta^{(i,t)}$ and $\theta^{(j,t-1)}$. Then, instead of choosing the perturbation kernel $K_t$ that maximizes equation~\eqref{eq:Qterm} over all possible kernels, the space of possible kernels is often restrained to a parametric family from which it is easy to sample and perform optimization. Different probability models may be used, with the most common being the uniform and the Gaussian distributions \citep{Sisson:2007aa,Toni:2009p9197,Liepe:2010eg}. In the following, we outline different classes of Gaussian perturbation kernels and compare their efficiency in terms of acceptance rates and computational cost.

\subsection{Component-wise perturbation kernel}\label{sec:componentwise}
 In most cases the particle is moved component-wise: for each component $1\leq j\leq d$ of the parameter vector $\theta=(\theta_1,\cdots\theta_d)$, $\theta_j$ is perturbed independently according to a Gaussian distribution with mean $\theta_j$ and variance $\sigma_j^2$. The parameters $\{\sigma_j\}_{1\leq j\leq d}$ may be fixed in advance, but more frequently \citep{Beaumont:2009be,Mckinley:2009dw,Toni:2010kt,Jasra:2010hg,Barnes:2011ua,Didelot:2010wo} adaptively chosen kernel widths, $\{\sigma_j^{(t)}\}_{1\leq j\leq d}$, are used which depend on the previous population --- the scale or variance is then indexed by the population index, $t$.
% For the perturbation kernel based on the uniform distribution, called in the following \textit{uniform kernel}, $\sigma_j^{(t)}$ is often set to the scale of the previous population, that is
%$$
%\sigma_j^{(t)}=\frac{1}{2}\left(\max_{1\leq k\leq N}\{\theta ^{(k,t)}_j\}-\min_{1\leq k\leq N}\{\theta ^{(k,t)}_j\}\right)\;.
%$$
Considering a kernel of the form
\be
K_t(\theta^{(t)}|\theta^{(t-1)})=\prod_{j=1}^d\frac{1}{\sqrt{2\pi}\sigma_j^{(t)}} \exp\left\{ -\frac{(\theta_j^{(t)}-\theta_j^{(t-1)})^2}{2\sigma_j^{(t)\; 2}}\right\}
\ee
and maximizing $Q(K_t,0,0,x)$ --- or more precisely $Q(K_t,\epsilon_{t-1},\epsilon_{t-1},x)$ --- \cite{Beaumont:2009be} showed that the optimal value of $\sigma_j^{(t)}$ is twice the empirical variance of the $j$-th component of the parameter vector in the previous population.
 %$$
 %\sigma_j^{(t)}=2 \var(\{\theta^{(k,t-1)}_j\}_{1\leq k\leq N})\;.
 %$$
Maximizing $Q(K_t,\epsilon_{t-1},\epsilon_{t},x)$, however, leads to a slightly different choice of $\sigma_j^{(t)}$ which is
%\be
%\sigma_j^{(t)}=\left(E_{p_{\epsilon_{t-1}}(\cdot|x)p_{\epsilon_{t}}(\cdot|x)}\left[ %(\theta_j^{(t)}-\theta_j^{(t-1)})^2\right]\right)^{1/2}\; .
%\ee
\be
\sigma_j^{(t)}=\left(\iint p_{\epsilon_{t-1}}(\cdot|x)p_{\epsilon_{t}}(\cdot|x) (\theta_j^{(t)}-\theta_j^{(t-1)})^2d\theta_j^{(t)}d\theta_j^{(t-1)}\right)^{1/2}\; .
\ee

This quantity can be approximated from the set $\{\theta^{(i,t-1)},\omega^{(i,t-1)},y^{(i,t-1)}\}_{1\leq i\leq N}$ as follows
\begin{equation}
\sigma_j^{(t)}\approx\left(\sum_{i=1}^N\sum_{k=1}^{N_0}\omega^{(i,t-1)}\tilde{\omega}^{(k,t-1)}(\tilde\theta_j^{(k,t-1)}-\theta_j^{(i,t-1)})^2\right)^{1/2}\; .\label{eq:componentwise}
\end{equation}
where
\begin{equation}
\left\{\left(\tilde\theta^{(k,t-1)},\tilde\omega^{(k,t-1)}\right)\right\}_{1\leq k\leq N_0}=\left\{\left(\theta^{(i,t-1)},\frac{\omega^{(i,t-1)}}{\bar{\omega}}\right)\text{ s.t. } \Delta(x, y^{(i,t-1)})\leq \epsilon_t, \; 1\leq i\leq N\right\}
\label{eq:thetatilde}
\end{equation}
and $\bar{\omega}$ is a normalizing constant such that 
%We denote by $\{\tilde\theta^{(k)}\}_{1\leq k\leq N_0}$ the set of particles $\left\{\theta_j^{(i,t-1)}\text{ s.t. } \Delta(x, y^{(i,t-1)})\leq \epsilon_t, \; 1\leq i\leq N\right\}$ and by $\{\tilde\omega^{(k)}\}_{1\leq k\leq N_0}$ the associated weights normalized such that 
$\sum_{k=1}^{N_0} \tilde\omega^{(k,t-1)}=1$.
\par
Another commonly used component-wise kernel is the  \textit{uniform kernel}, which consists of perturbing the $j$-th component of particle $\theta$ to any value in the interval $\left[\theta_j-\sigma_j^{(t)};\theta_j-\sigma_j^{(t)}\right]$ with density $1/2\sigma_j^{(t)}$. 
A natural choice is to set the parameter $\sigma_j^{(t)}$ to the scale of the previous population, that is
$$
\sigma_j^{(t)}\approx\frac{1}{2}\left(\max_{1\leq k\leq N}\{\theta ^{(k,t-1)}_j\}-\min_{1\leq k\leq N}\{\theta^{(k,t-1)}_j\}\right)\;.
$$
Note that the main difference between the uniform and the component-wise normal kernels concerns their support: bounded vs. unbounded. 

\subsection{Multivariate normal perturbation kernels}\label{sec:multivariateNormal}
\begin{figure}
\centering
\subfigure{
\begin{tikzpicture}[scale=0.48]
\begin{pgfonlayer}{background}
\node at (0,0)  { 
\includegraphics[width=0.3\textwidth]{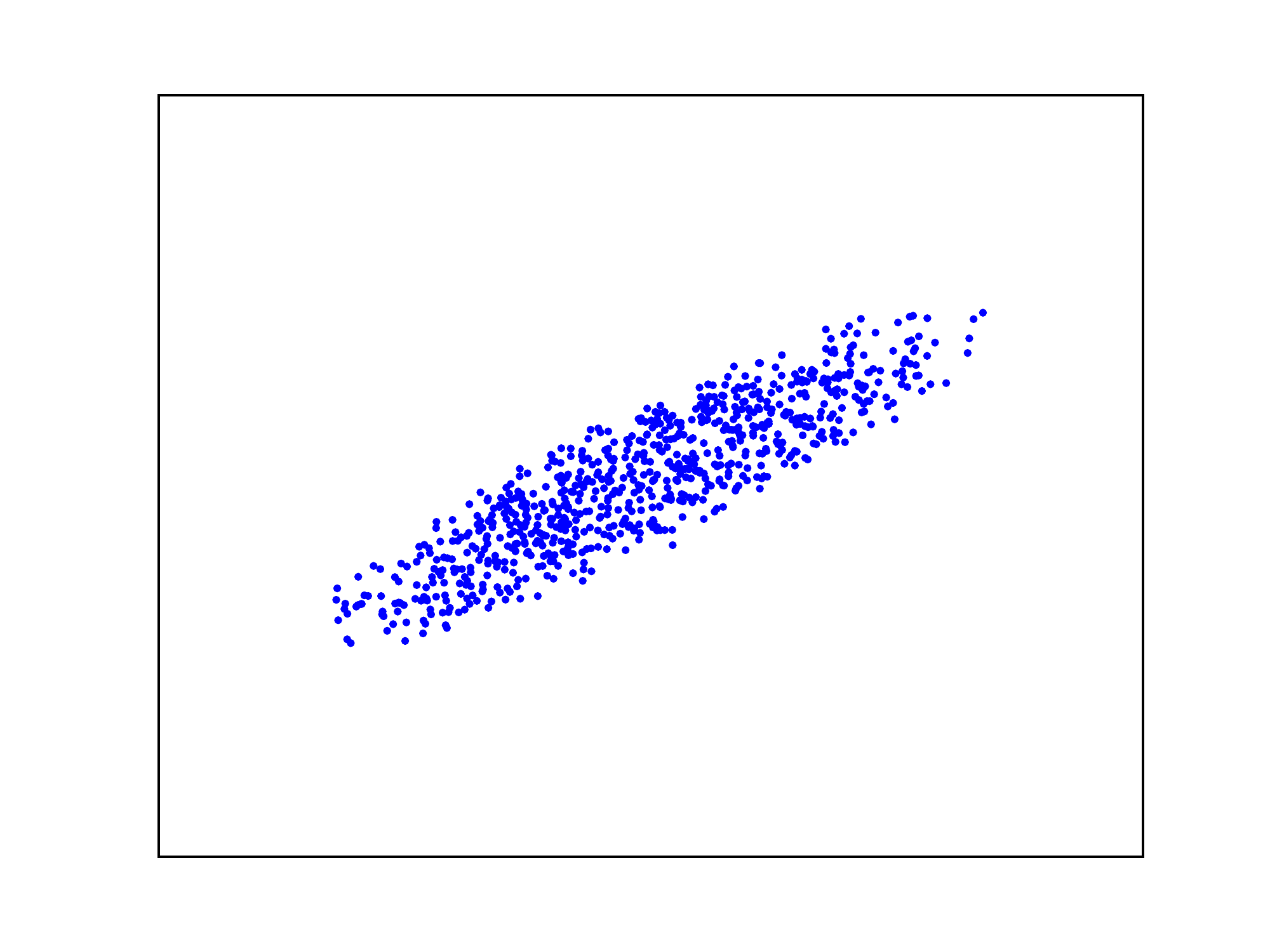} 
};
\end{pgfonlayer}
\draw[draw=blue,rotate around={25:(0,-0.05)}] (0,-0.05) ellipse (3.5 and 0.6);
\draw[line width=0.6,draw=red] (-3.7,-2.2) rectangle (2.7,1.1);
\draw (4,-3.6) node [black] { $\theta_1$};
\draw (-4.6,3) node [black] { $\theta_2$};
\draw[fill=red,draw=red] (-0.5,-0.5) circle(0.06);
\end{tikzpicture}
%\caption{A population of particles and the support of a uniform perturbation kernel around one particule (red point).}
%\label{fig:ellipseUniform}
}
\hspace{-0.8cm}
\subfigure{
\begin{tikzpicture}[scale=0.48]
\begin{pgfonlayer}{background}
\node at (0,0)  { 
\includegraphics[width=0.3\textwidth]{fig/ellipse.pdf} 
};
\end{pgfonlayer}
\draw[draw=blue,rotate around={25:(0,-0.05)}] (0,-0.05) ellipse (3.5 and 0.6);
\draw[line width=0.6,draw=red,rotate around={0:(-0.5,-0.5)}] (-0.5,-0.5) ellipse (3.4 and 2);
\draw[line width=0.6,,draw=red,rotate around={0:(-0.5,-0.5)}] (-0.5,-0.5) ellipse (3 and 1.6);
\draw[line width=0.6,,draw=red,rotate around={0:(-0.5,-0.5)}] (-0.5,-0.5) ellipse (2.6 and 1.39);
\draw[line width=0.6,,draw=red,rotate around={0:(-0.5,-0.5)}] (-0.5,-0.5) ellipse (2 and 1.07);
\draw[line width=0.6,,draw=red,rotate around={0:(-0.5,-0.5)}] (-0.5,-0.5) ellipse (1.5 and 0.8);
\draw (4,-3.6) node [black] { $\theta_1$};
\draw (-4.6,3) node [black] { $\theta_2$};
\draw[fill=red,draw=red] (-0.5,-0.5) circle(0.06);
\end{tikzpicture}
}
\hspace{-0.8cm}
\subfigure{
\begin{tikzpicture}[scale=0.48]
\begin{pgfonlayer}{background}
\node at (0,0)  { 
\includegraphics[width=0.3\textwidth]{fig/ellipse.pdf} 
};
\end{pgfonlayer}
\draw[line width=0.6,,draw=blue,rotate around={25:(0,-0.05)}] (0,-0.05) ellipse (3.5 and 0.6);
\draw[line width=0.6,,draw=red,rotate around={25:(-0.5,-0.5)}] (-0.5,-0.5) ellipse (3.4 and 0.5);
\draw[line width=0.6,,draw=red,rotate around={25:(-0.5,-0.5)}] (-0.5,-0.5) ellipse (3 and 0.4);
\draw[line width=0.6,,draw=red,rotate around={25:(-0.5,-0.5)}] (-0.5,-0.5) ellipse (3.4 and 0.5);
\draw[line width=0.6,,draw=red,rotate around={25:(-0.5,-0.5)}] (-0.5,-0.5) ellipse (2.6 and 0.3);
\draw[line width=0.6,,draw=red,rotate around={25:(-0.5,-0.5)}] (-0.5,-0.5) ellipse (2 and 0.2);
\draw (4,-3.6) node [black] { $\theta_1$};
\draw (-4.6,3) node [black] { $\theta_2$};
\draw[fill=red,draw=red] (-0.5,-0.5) circle(0.06);
\end{tikzpicture}
}
\caption{A population of particles and isodensity curves for a uniform kernel (left), a component-wise normal kernel (centre) and a multivariate normal perturbation kernel (right) around one particle (red point).}
\label{fig:ellipse}
\end{figure}

Consider a population of two-dimensional parameters that are highly correlated. The perturbation of a particle according to the uniform kernel consists in sampling a parameter uniformly in a rectangle whose sides are parallel to the axes (see Figure \ref{fig:ellipse} left). Similarly, the density levels of the component-wise normal kernel are ellipsoids whose principal axes are parallel to the parameter axes (see Figure \ref{fig:ellipse} centre). For highly correlated parameters the use of component-wise perturbation kernels in the ABC SMC framework can lead to a small acceptance rate because they can inadequately reflect the structure of the true posterior. 

Instead of using a component-wise kernel it may thus be more efficient to take into account the correlations between the different elements of the parameter vectors, in effect perturbing the particles according to a multivariate normal distribution with a covariance matrix $\Sigma^{(t)}$, which depends on the covariance of the previous population. Figure \ref{fig:ellipse} (right) represents a \textit{multivariate normal perturbation kernel} for a matrix $\Sigma^{(t)}$ proportional to the covariance of the previous population. We observe that fewer particle proposals are likely to be rejected with this perturbation kernel compared to the uniform or component-wise normal one.
\par
The multivariate normal perturbation kernel relies on the covariance matrix $\Sigma^{(t)}$ which depends on the previous population. As before, it is possible to calculate the optimal covariance matrix $\Sigma^{(t)}$ using the Kullback-Leibler divergence minimization approach (see also \citep{Cappe:2008ht}). Maximizing equation~\eqref{eq:Qterm} for  
\[K_t(\theta^{(t)}|\theta^{(t-1)})=(2\pi)^{-d/2}\left(\det\Sigma^{(t)}\right)^{-1/2}\exp\left\{-\frac{1}{2}\left(\theta^{(t)}-\theta^{(t-1)}\right)^T \left(\Sigma^{(t)}\right)^{-1}\left(\theta^{(t)}-\theta^{(t-1)}\right)\right\}\]
with respect to $\Sigma^{(t)}$
leads to maximizing the real-valued function 
%$$
%g(M)=\log\det\left(M\right)- E_{p_{\epsilon_{t-1}}(\cdot|x)p_{\epsilon_{t}}(\cdot|x)}\left[(\theta^{(t)}-\theta^{(t-1)})^T M(\theta^{(t)}-\theta^{(t-1)})\right]\;,
%$$
$$
g(M)=\log\det\left(M\right)- \iint p_{\epsilon_{t-1}}(\theta^{(t-1)}|x)p_{\epsilon_{t}}(\theta^{(t)}|x)(\theta^{(t)}-\theta^{(t-1)})^T M(\theta^{(t)}-\theta^{(t-1)})d\theta^{(t)}d\theta^{(t-1)}\;,
$$
with respect to the symmetric $d\times d$ matrix $M$, and defining $\Sigma^{(t)}=M^{-1}$. We denote by $v^T$ the transpose of vector $v$. 
%Taking the partial derivatives of the function $g: \Rset^{d\times d}\to \Rset $, we obtain, for all %$1\leq i,j\leq d$,
%\begin{align*}
 %\frac{\partial g(M)}{\partial M_{ij}}&=(M^{-1})_{ij}-\frac{\partial}{\partial %M_{ij}}E_{p_{\epsilon_{t-1}}(\cdot|x)p_{\epsilon_{t}}(\cdot|%x)}\left[\tr\left((\theta^{(t)}-\theta^{(t-1)})^TM(\theta^{(t)}-\theta^{(t-1)})\right)\right]\\
%&=(M^{-1})_{ij}-\frac{\partial}{\partial M_{ij}} \tr\left(MC\right)=(M^{-1})_{ij}-%C_{ji}=(M^{-1})_{ij}-C_{ij}\;.
%\end{align*}
%where $\tr(M)$ denotes the trace of the matrix $M$, and the symmetric matrix $C$ is equal to 
%$$
%C=E_{p_{\epsilon_{t-1}}(\cdot|x) p_{\epsilon_{t}}(\cdot|x)}\left[(\theta^{(t)}-\theta^{(t-1)})%(\theta^{(t)}-\theta^{(t-1)})^T\right]\;.
%$$ 
We then obtain that the covariance matrix $\Sigma^{(t)}$ of the optimal kernel in the multivariate Gaussian family is 
%$$
%\Sigma^{(t)}=E_{p_{\epsilon_{t-1}}(\cdot|x) p_{\epsilon_{t}}(\cdot|x)}\left[(\theta^{(t)}-\theta^{(t-1)})(\theta^{(t)}-\theta^{(t-1)})^T\right]\;.
%$$
$$
\Sigma^{(t)}=\iint p_{\epsilon_{t-1}}(\theta^{(t-1)}|x)p_{\epsilon_{t}}(\theta^{(t)}|x)(\theta^{(t)}-\theta^{(t-1)})(\theta^{(t)}-\theta^{(t-1)})^T d\theta^{(t)}d\theta^{(t-1)}\;.
$$
Proceeding in a similar fashion to the component-wise case, if we assume that $\epsilon_t=\epsilon_{t-1}$, then the optimal covariance matrix can be approximated by twice the empirical covariance matrix of the population $t-1$. In the general case, however, an optimal choice of the covariance matrix $\Sigma^{(t)}$ for the multivariate normal perturbation kernel is then approximated by
\be
\Sigma^{(t)}\approx \sum_{i=1}^N\sum_{k=1}^{N_0}\omega^{(i,t-1)}\tilde{\omega}^{(k)}(\tilde\theta^{(k)}-\theta^{(i,t-1)}) (\tilde\theta^{(k)}-\theta^{(i,t-1)})^T \;.
\label{eq:covPop}
\ee
where, $\{\tilde\theta^{(k)}\}_{1\leq k\leq N_0}$ and $\{\tilde\omega^{(k)}\}_{1\leq k\leq N_0}$ are defined by equation~\eqref{eq:thetatilde}. In the following, if nothing else is specified, the {\it multivariate normal kernel} refers to the kernel with this choice of covariance matrix. This result generalises the work of \cite{Beaumont:2009be}.
%: their optimality criteria to cases where either only a single parameter is to be estimated or where the covariance matrix is diagonal. Our scheme generalises their criterion to a much broader class of kernels.

\subsection{Local perturbation kernels} 

In many applications the parameters of the system are highly correlated but in a non-linear way.  The multivariate normal and the component-wise normal kernels discussed above may then behave similarly (see for example the toy model described in section~\ref{sec:banana}). Indeed, the covariance matrix based on all the previous particles yields only limited information about the local correlation among the individual components of the parameter vectors. In such cases it is interesting to consider the use of a local covariance matrix which now may differ between particles. In the following we discuss three local perturbation kernels for which each particle $\theta$ is perturbed according to a multivariate normal kernel whose covariance matrix $\Sigma^{(t)}_\theta$ is a function of $\theta$. 

\subsubsection{The multivariate normal kernel with $M$ nearest neighbours}
\label{sec:kNN}
The \textit{multivariate normal kernel with $M$ neighbours} follows this principle: for each particle $\theta\in\{\theta^{(l,t-1)},1\leq l\leq N\}$, the $M$-nearest neighbours of $\theta$ are selected, and the perturbed particle is sampled according to a multivariate normal distribution of mean $\theta$ and of covariance the empirical covariance $\Sigma_{\theta,M}^{(t)}$ based on the $M$ nearest neighbours of  $\theta$.

The main drawback of this perturbation kernel is that the parameter $M$ typically needs to be fixed in advance before any of the intricacies of the posterior are known. 
%Figures \ref{fig:boomerang} (centre) and \ref{fig:boomerang} (right) illustrate the effect of the parameter $M$ on the perturbation kernel. 
Using too small a value may lead to a lack of exploration of parameter space, while too large a value of $M$ would offer little or no advantage compared to the standard multivariate normal kernel. Ideally, a mixture of multivariate normal kernels with different values of $M$ could be used; however, in practice, this solution is computationally too expensive. 

\subsubsection{The multivariate normal kernel with optimal local covariance matrix}
\label{sec:OLCM}
The theoretical calculation of the optimal covariance matrix above (see Section~\ref{sec:multivariateNormal}) may be adapted to identify an optimal local covariance matrix. Let's consider a particle $\theta^{(t-1)}$ which has been sampled from the previous population $\{\theta^{(k,t-1)}\}_{1\leq k\leq N}$. To determine the covariance matrix $\Sigma_{\theta^{(t-1)}}^{(t)}$ of a multi-variate normal pertubation kernel centered in $\theta^{(t-1)}$ we derive a criterion $Q(K_t,\epsilon_t,x)$ similar to the criterion defined in equation~\eqref{eq:Qterm} expect that the particle $\theta^{(t-1)}$ is fixed:
\be
Q(K_t,\epsilon_t,x)=\int p_{\epsilon_t}(\theta^{(t)}|x) \log
	K_t(\theta^{(t)}|\theta^{(t-1)}) d\theta^{(t)}	
\ee
which is equal to

%\be
%E_{p_{\epsilon_{t-1}}(\cdot|x)p_{\epsilon_{t}}(\cdot|x)}\left[\log\det  \left(\Sigma_{\theta^{(t-1)}}^{(t)}\right)^{-1}-\frac{d}{2}\log(2\pi)-\frac{1}{2}(\theta^{(t)}-\theta^{(t-1)})^T\left(\Sigma_{\theta^{(t-1)}}^{(t)}\right)^{-1} (\theta^{(t)}-\theta^{(t-1)})\right]
%\ee
\be
\int p_{\epsilon_{t}}(\cdot|x)\log\det  \left(\Sigma_{\theta^{(t-1)}}^{(t)}\right)^{-1}-\frac{d}{2}\log(2\pi)-\frac{1}{2}(\theta^{(t)}-\theta^{(t-1)})^T\left(\Sigma_{\theta^{(t-1)}}^{(t)}\right)^{-1} (\theta^{(t)}-\theta^{(t-1)})d\theta^{(t)}
\ee
when $K_t(\theta^{(t)}|\theta^{(t-1)})$ is a multi-variate normal kernel centred on $\theta^{(t-1)}$ with covariance matrix $\Sigma_{\theta^{(t-1)}}^{(t)}$. Maximizing the equation above with respect to $\Sigma_{\theta^{(t-1)}}^{(t)}$ the optimal local covariance matrix is
%$$
%E_{p_{\epsilon_{t-1}}(\cdot|x)}\left[ \Sigma_{\theta^{(t-1)}}^{(t)}- E_{p_{\epsilon_{t}}(\cdot|x)}\left[(\theta^{(t)}-\theta^{(t-1)})(\theta^{(t)}-\theta^{(t-1)})^T\right]\right]=0\;.
%$$
$$
\Sigma_{\theta^{(t-1)}}^{(t)}=\int p_{\epsilon_{t}}(\theta^{(t)}|x)(\theta^{(t)}-\theta^{(t-1)})(\theta^{(t)}-\theta^{(t-1)})^Td\theta^{(t)}\;,
$$
%In particular, the set of covariance matrices such that, for all $t$ and each particle %$\theta^{(j,t-1)}$, $1\leq j\leq N$, 
%$$
%\Sigma_{\theta^{(j,t-1)}}^{(t)}=E_{p_{\epsilon_{t}}(\cdot|x)}\left[(\theta^{(t)}-\theta^{(j,t-1)})(\theta^{(t)}-\theta^{(j,t-1)})^T\right]
%$$ 
%satisfies the above condition. 
which can be approximated from the set $\left\{\theta^{(i,t-1)},\omega^{(i,t-1)}, y^{(i,t-1)}\right\}_{1\leq i\leq N}$ as follows
$$
\Sigma_{\theta^{(t-1)}}^{(t)}\approx\sum_{k=1}^{N_0} \tilde{\omega}^{(k)}(\tilde{\theta}^{(k)}-\theta^{(t-1)})(\tilde{\theta}^{(k)}-\theta^{(t-1)})^T\;,
$$
where, $\{\tilde\theta^{(k)}\}_{1\leq k\leq N_0}$ and $\{\tilde\omega^{(k)}\}_{1\leq k\leq N_0}$ are defined by equation~\eqref{eq:thetatilde}. To better understand the covariance matrix  $\Sigma_{\theta^{(t-1)}}^{(t)}$ we remark, by a classical bias-variance decomposition, that it is equal to the covariance of the particles from the previous population corresponding to distances smaller than $\epsilon_t$ (i.e. the weighted particle denoted $\{\tilde\theta^{(k)},\tilde\omega^{(k)}\}_{1\leq k\leq N_0}$ through this article) plus a bias term related to the discrepancy between the mean of this population and the particle of interest $\theta^{(t-1)}$:
$$
\Sigma_{\theta^{(t-1)}}^{(t)}\approx \sum_{k=1}^{N_0} \tilde{\omega}^{(k)}(\tilde{\theta}^{(k)}-m)(\tilde{\theta}^{(k)}-m)^T+ (m-\theta^{(t-1)})(m-\theta^{(t-1)})^T
$$
where $m=\sum_{k=1}^{N_0} \tilde{\omega}^{(k)}\tilde{\theta}^{(k)}$.
\par
The multivariate normal perturbation kernel with a covariance matrix as defined above will be referred to as the \textit{multivariate normal kernel with OLCM} (where OLCM stands for optimal local covariance matrix). We now have a different covariance matrix $\Sigma_{\theta^{(j,t-1)}}^{(t)}$ for each particle $\theta^{(j,t-1)}$ of the previous population.

\subsubsection{Perturbation kernel based on the Fisher information for model defined by ordinary or stochastic differential equations}
Often in molecular biology the time evolution of species abundance is modelled by a system of ordinary differential equations (ODE), stochastic differential equations (SDE) or a chemical master equation (CME). For these types of models it is possible to use information from the generative model --- via the Fisher Information Matrix \citep{Rao:45,Mackay:2003aa,Cox2006} --- to compute a local covariance matrix for the perturbation kernel. The Fisher Information Matrix (FIM) defined as 
$$
I(\theta)=-E_X\left[\frac{\partial^2 }{\partial\theta^2}\log f(X|\theta)\right]
$$ 
measures the amount of information that the observable random variable $X$ carries about the parameter $\theta$. 
As previously mentioned, the ABC algorithm is mainly used when the likelihood function $f(\cdot|\theta)$ is not known, and so the Fisher Information Matrix can not be computed exactly. 
Nevertheless, \cite{Komorowski:2011cn} have developed a method  that approximates the FIM for deterministic and stochastic dynamical systems represented by ODEs and by SDEs (using the linear noise approximation). This evaluation of the FIM can be applied as part of the ABC SMC procedure progresses and so the perturbation kernels can adapt appropriately. Despite the fact that the following kernels cannot be computed in the general case, we consider them here because of their potential use for inference in dynamical systems.
\par
\begin{figure}
\centering
%\subfigure{
%\begin{tikzpicture}[scale=0.75]
%\begin{pgfonlayer}{background}
%\node at (0,0)  { 
%\includegraphics[width=0.48\textwidth]{fig/boomerang.pdf} 
%};
%\end{pgfonlayer}
%\draw[->,line width=0.8,red] (1.3,0.7) -- (4.25,-1);
%\draw[->,line width=0.8,red] (1.3,0.7) -- (0.3,0);
%\draw (4,-3.6) node [black] { $\theta_1$};
%\draw (-4.6,3) node [black] { $\theta_2$};
%\draw[fill=red,draw=red] (1.3,0.7) circle(0.06);
%\end{tikzpicture}
%}
%\hspace{-0.8cm}
%\subfigure{
%\begin{tikzpicture}[scale=0.75]
%\begin{pgfonlayer}{background}
%\node at (0,0)  { 
%\includegraphics[width=0.48\textwidth]{fig/boomerang.pdf} 
%};
%\end{pgfonlayer}
%\draw[->,line width=0.8,red] (2.2,-0.5) -- (0.2,-3.2);
%\draw[->,line width=0.8,red] (2.2,-0.5) -- (1.8,-0.2);
%\draw (4,-3.6) node [black] { $\theta_1$};
%\draw (-4.6,3) node [black] { $\theta_2$};
%\draw[fill=red,draw=red] (2.2,-0.5) circle(0.06);
%\end{tikzpicture}
%}
\includegraphics[width=\textwidth]{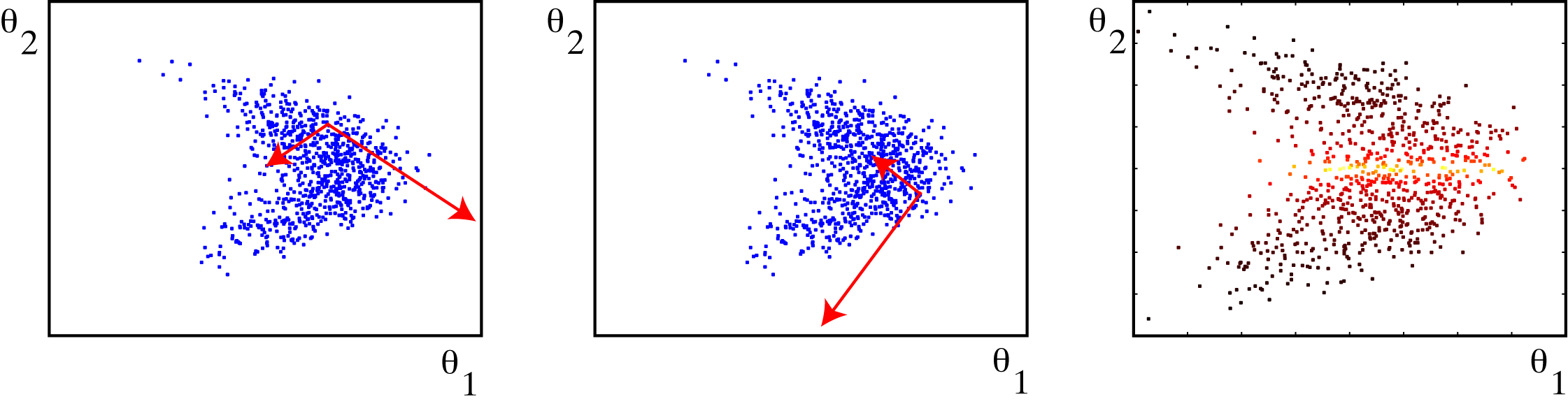} 
%}
\caption{Local kernel based on the Fisher Information Matrix $I(\theta)$. Left and centre: The eigenvectors of $I^{-1}(\theta)$ (red arrows) of size proportional to the eigenvalues for two different particles $\theta$. Right: Logarithm of the determinant of $I^{-1}(\theta)$ for each particle $\theta$. The maximum value of $\det(I^{-1}(\theta))$ over the population is equal to $85745$ (yellow points) and the minimum one is equal to $0.14$ (black points).}
\label{fig:ev}
\end{figure}

In the Laplace expansion the eigenvectors and eigenvalues of the inverse of the FIM $I(\theta)$ map out ellipsoidal levels of equal density around the parameter $\theta$. This immediately suggests the use of the matrix $I^{-1}(\theta)$ as the covariance matrix for a multivariate normal perturbation kernel. 
%The eigenvectors of $I^{-1}(\theta)$ (scaled such that their lengths are proportional to their corresponding eigenvalues) are represented for two different values of $\theta$ in Figure \ref{fig:ev}. 
The directions of the eigenvectors of $I^{-1}(\theta)$ and the relative size of their eigenvalues are indeed both relevant for perturbing the parameter $\theta$ efficiently (see Figure~\ref{fig:ev}).
% However, in Figure \ref{fig:detFIM} we observe that the determinant of $I^{-1}(\theta)$ varies exponentially with $\theta$ over $5$ orders of magnitude. 
However, the determinant of $I(\theta)$ is a measure of the amount of information available around $\theta$ and may vary exponentially with $\theta$ (see Figure \ref{fig:ev} Right); its value is very small for some parameters $\theta$ and this leads to a perturbation kernel with too large a covariance. On the other hand, if the determinant of $I(\theta)$ is large, additional information may be gained by moving only in the direct vicinity of $\theta$, and a perturbation kernel based on the inverse of the FIM explores only the immediate neighbourhood of the parameter. 

We therefore propose scaling the matrix $I^{-1}(\theta)$ such that its determinant varies in a more controlled manner. We consider two versions of the \textit{multivariate normal perturbation kernel based on the FIM}: the first one consists of normalizing the matrix such that its determinant is equal to the determinant of the empirical covariance matrix of the previous population
$$
\Sigma_{\theta}^{(t)}=\left(\frac{\det(\Sigma^{(t)})}{\det(I^{-1}(\theta))}\right)^{1/d} I^{-1}(\theta),
$$
where $\Sigma^{(t)}$ is the matrix defined in equation~\eqref{eq:covPop}; the second approach consists of normalizing the matrix such that its determinant is equal to the determinant of the empirical covariance matrix $\Sigma_{\theta,M}^{(t)}$ based on the $M$ nearest neighbours of the particle defined according to
$$
\Sigma_{\theta,M}^{(t)}=\left(\frac{\det(\Sigma^{(t)}_{\theta,M}))}{\det(I^{-1}(\theta))}\right)^{1/d} I^{-1}(\theta)\;.
$$
The parameter $M$ may, for instance, be equal to 20\% of the previous population.

\section{Numerical results}
We first apply the ABC SMC algorithm with different kernels to three illustrative examples, which exhibit certain pathological features that highlight the differences between the perturbation kernels considered here. In order to analyze the impact of the kernel choice rather than any extraneous factors, we fix the threshold schedule $\{\epsilon_t\}_t$ to $(160,120,80,60,40,30,20,15,10,8,6,4,3,2,1)$ and use the same for every simulation.  However, in practice we strongly advise using an adaptive choice of the $\epsilon$ threshold. All the simulation have been done using the software \textit{ABC-SysBio} \citep{Liepe:2010eg} version 1.03 (which allows for adaptive kernels, too) using an euclidian distance to compare simulated and observed data.
\subsection{Ellipsoid shape}
\begin{figure}
\centering
\includegraphics[width=0.8\textwidth]{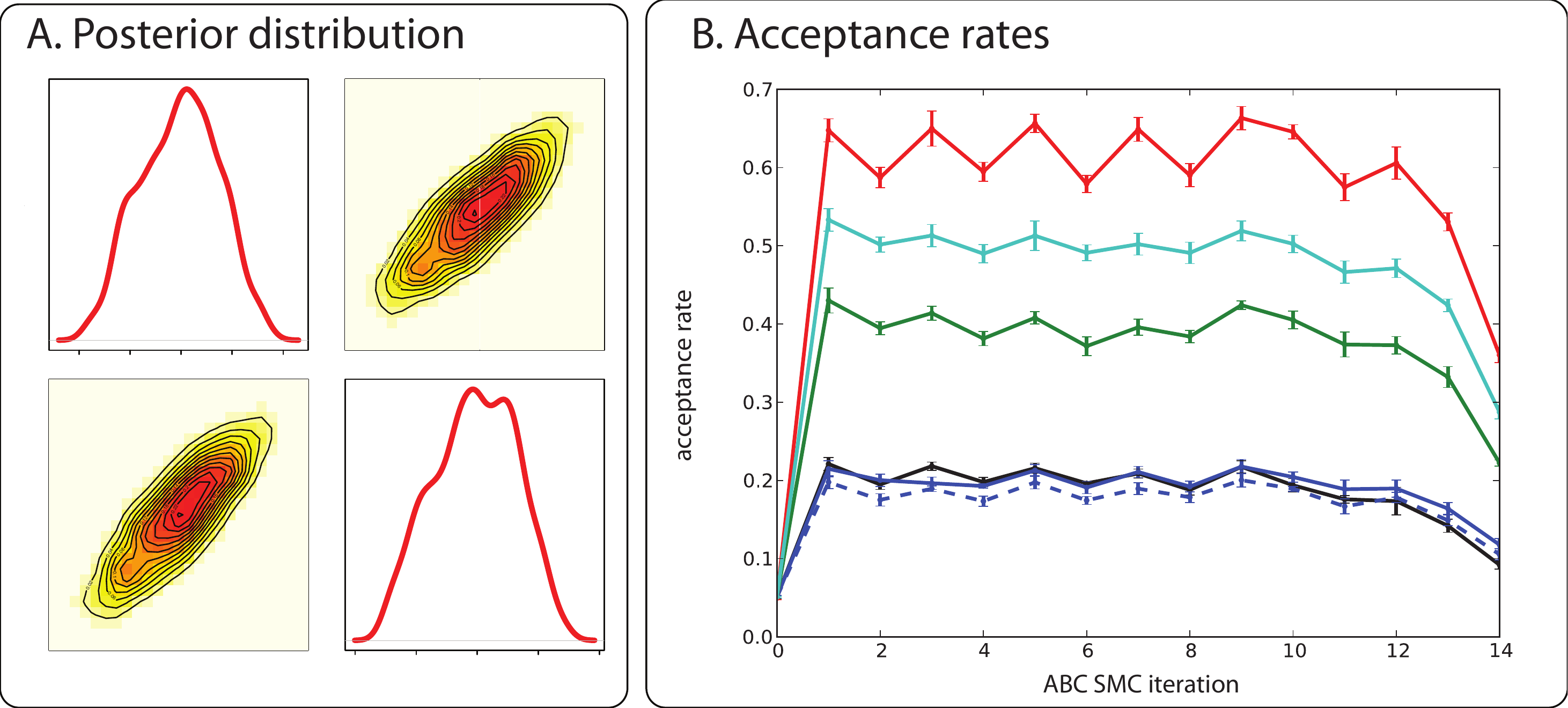}
\caption{A. Posterior distribution for an ellipsoid posterior. B. Average of the acceptance rate over 10 independent runs for $6$ different kernels: the uniform kernel (green), the component-wise normal kernel (blue), the component-wise normal kernel proposed by \cite{Beaumont:2009be} (dashed blue), the multivariate normal kernel (black), the multivariate normal kernel with $50$ neighbours (red), the multivariate normal kernel with OLCM (cyan). }
\label{fig:ellipsoid}
\end{figure}
We begin with a toy example where the prior distribution of the two dimensional parameter is a uniform distribution on the square $[-50,50]\times [-50,50]$ and the likelihood function is given by
$$
x\sim\mathcal{N}\left((\theta_1-2\theta_2)^2+(\theta_2-4)^2,1\right)\; .
$$
It is assumed that $x=0$ is observed. The posterior density is then 
$$
p(\theta|x)\propto \phi(0;(\theta_1-2\theta_2)^2+(\theta_2-4)^2,1)\1_{[-50,50]\times [-50,50]}(\theta)\;
$$ 
where $\phi(x;\mu,\sigma^2)$ is the one dimensional normal density with mean $\mu$ and variance $\sigma^2$, and is represented in Figure~\ref{fig:ellipsoid} A. The ABC SMC algorithm is used to estimate $p_\epsilon(\theta|x)$ with $N=800$ particles. We compare $6$ different perturbation kernels: 
1) the uniform kernel (Section~\ref{sec:componentwise}), 
2) the component-wise normal kernel (Section~\ref{sec:componentwise}), 
3) the component-wise normal kernel proposed by \cite{Beaumont:2009be}, 
4) the multivariate normal kernel with the covariance matrix computed from the whole previous population (Section~\ref{sec:multivariateNormal}), 
5) the multivariate normal kernel whose covariance matrix is computed according to the $M$ nearest neighbours of each particle (Section~\ref{sec:kNN}), with $M=50$, and 
6) the multivariate normal kernel with OLCM (Section~\ref{sec:OLCM}).

We reiterate that the variance of the component-wise normal kernel proposed by \cite{Beaumont:2009be} is slightly different than the one we propose here: they suggest to use twice the empirical variance of the previous population as a variance whereas we take into account the new threshold $\epsilon_t$, as described in equation~\eqref{eq:componentwise}. 
\par

Figure \ref{fig:ellipsoid} B shows that the acceptance rate differs significantly between kernels. The uniform kernel has an acceptance rate roughly equal to that of the component-wise normal kernel. Moreover, the two versions of the component-wise normal kernels have similar acceptance rates, with a slightly better performance for the one taking into account the difference between the successive threshold values. Given the shape of the posterior distribution, it is easy to understand that a multivariate normal kernel results in a larger acceptance rate than the other kernels. Since the two components of the parameters are strongly correlated, using an estimate of the covariance from the previous population instead of an estimation of only the  component-wise variances makes a marked difference on the acceptance rates. Both the multivariate normal kernel based on the $50$ nearest neighbours and the one based on the OLCM result in acceptance rates over two times higher than the component-wise kernels.

\subsection{Ring shape}
\begin{figure}
\centering
 \includegraphics[width=0.80\textwidth]{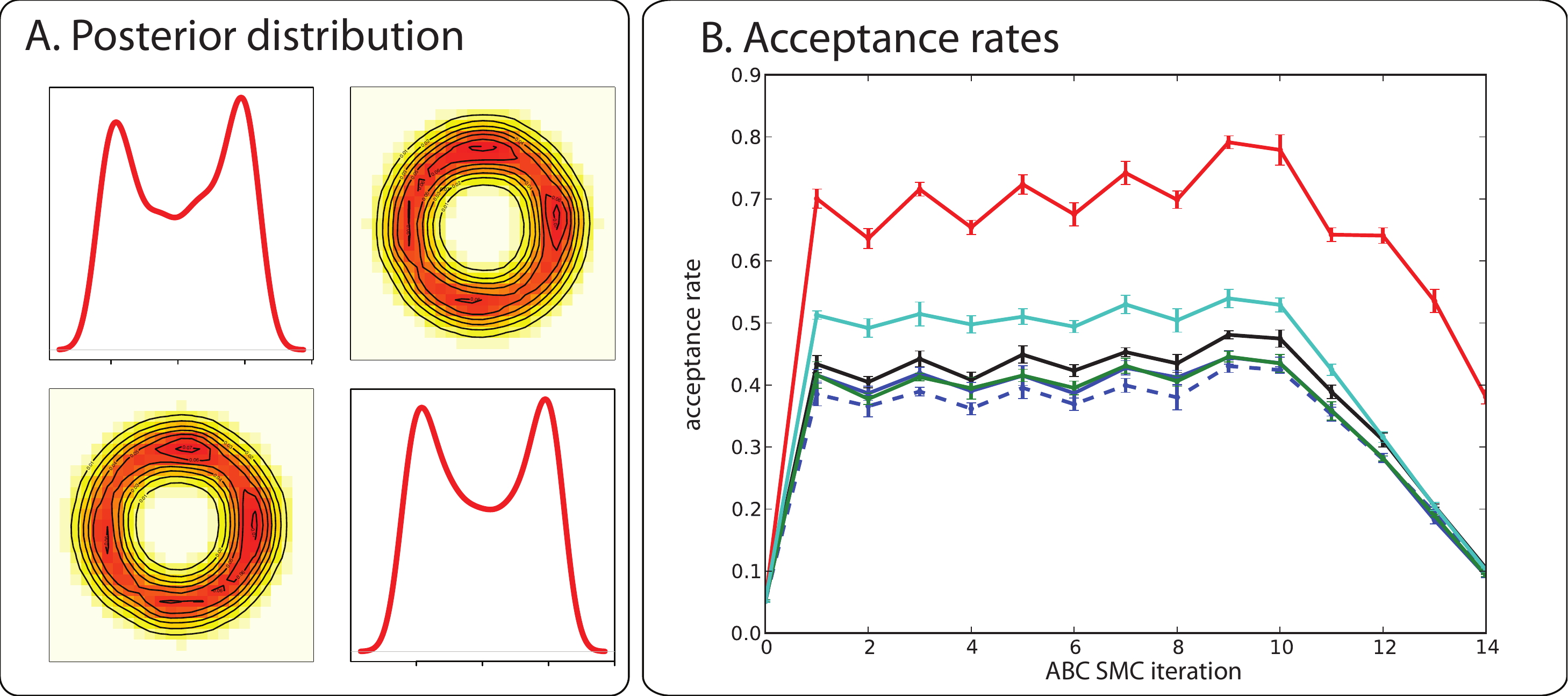}
\caption{A. Posterior distribution for an ellipsoid posterior. B. Average of the acceptance rate over 10 independent runs for $6$ different kernels: the uniform kernel (green), the component-wise normal kernel (blue), the component-wise normal kernel proposed by \cite{Beaumont:2009be} (dashed blue), the multivariate normal kernel (black), the multivariate normal kernel with $50$ neighbours (red), the multivariate normal kernel with OLCM (cyan). }
\label{fig:ring}
\end{figure}

In the second toy example, the prior distribution of the two dimensional parameter is still a uniform distribution on the square $[-50,50]\times [-50,50]$ but the likelihood function is now given by
$$
x\sim\mathcal{N}\left(\theta_1^2+\theta_2^2,0.5\right)\;.
$$
 Again we assume that $x=0$ is observed; the posterior density is then 
$$
p(\theta|x)\propto \phi(0;\theta_1^2+\theta_2^2,0.5)\1_{[-50,50]\times [-50,50]}(\theta)\;.
$$ 
As in the previous example we used the ABC SMC algorithm with $N=800$ particles and compare the same $5$ perturbation kernels. 
\par

The posterior distribution, represented by Figure \ref{fig:ring} A, has a ring shape centred around $0$. In this case, in contrast to the previous example, the multivariate normal perturbation kernel using an estimate of the covariance based on the previous population, as well as the OLCM version of it, have an acceptance rate similar to the component-wise normal perturbation kernel. In this example the correlation between the two parameters, $\theta_1$ and $\theta_2$, at the whole population level is weak. This kind of shape requires a more local perturbation kernel in order to obtain higher acceptance rates. This is the case for the perturbation kernel based on the covariance matrix computed from the $50$ nearest neighbours.

\subsection{Banana shape}\label{sec:banana}
\begin{figure}
\centering
 \includegraphics[width=0.80\textwidth]{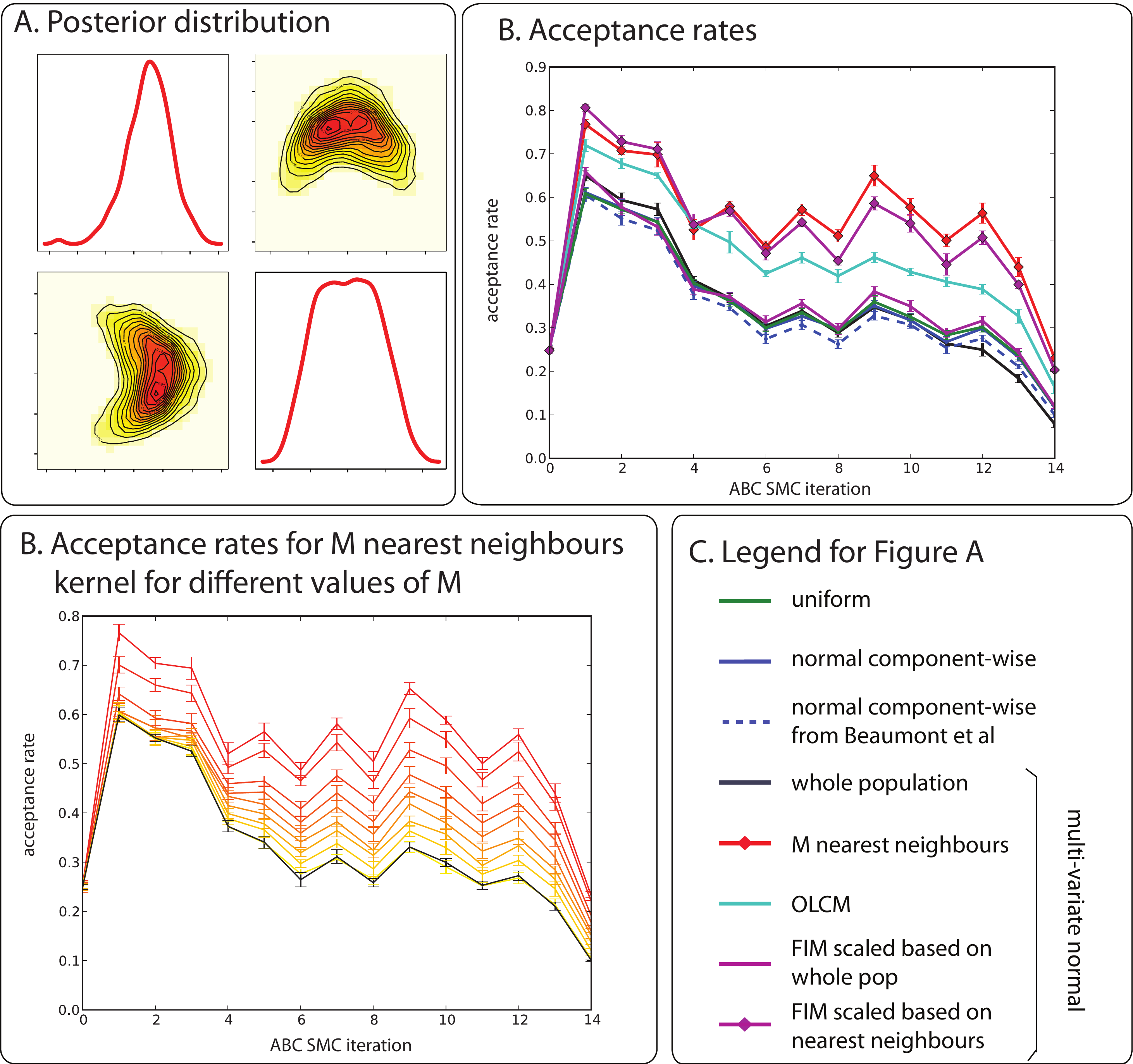}
\caption{A. Posterior distribution for an ellipsoid posterior. B. Average of the acceptance rate over 10 independent runs for $8$ different kernels. C. Acceptance rate for multivariate normal kernels based on the $M$ neighbours for $M\in\{50,100,200,300,400,500,600,700,800\}$ (from red to yellow) and the multivariate kernel with an estimated covariance based on the whole population (black). D. Legend for Figure B.}
\label{fig:banana}
\end{figure}
The third example we consider is one of the canonical examples of a posterior distribution which poses a challenge to simple kernels: the so-called `banana-shape' distribution in two dimensions \citep{haario1999adaptive}. The likelihood function is given by 
$$
\left(\begin{array}{c}x_1\\x_2\end{array}\right)\sim\mathcal{N}_2\left(\left(\begin{array}{c}\theta_1\\\theta_1+\theta_2^2\end{array}\right),\left( \begin{array}{cc}
1 & 0 \\
0 & 0.5 \end{array} \right)\right)
$$
and we use a uniform prior distribution on the square $[-50,50]\times [-50,50]$.
 It is assumed that $x=(0,0)$ is observed. The posterior density is then 
$$
p(\theta|x)\propto \Phi\left(\left(\begin{array}{c}0\\0\end{array}\right);\left(\begin{array}{c}\theta_1\\\theta_1+\theta_2^2\end{array}\right),\left( \begin{array}{cc}
1 & 0 \\
0 & 0.5 \end{array} \right)\right)\1_{[-50,50]\times [-50,50]}(\theta)
$$ 
where $\Phi(x;v,\Sigma)$ is the multi-dimensional normal density of mean $v$ and covariance $\Sigma$. 
We use the same ABC SMC settings and again compare the $6$ previous perturbation kernels, as well as two versions of the multivariate normal perturbation kernel where the covariance matrix is proportional to the inverse of the FIM. Here the FIM is exactly computable:
$$
I(\theta)=\left(\begin{array}{cc} 1.5& \theta_2\\ \theta_2 & 2\theta_2^2   \end{array} \right)\;.
$$
When $\theta_2=0$ we replace it by a very small value, $10^{-4}$, such that $I(\theta)$ is no longer singular; safeguarding against singular FIMs is straightforward and unproblematic and a sensible precaution when running such algorithms without manual intervention.
\par
The posterior distribution is represented in Figure \ref{fig:banana} A.  As in the ring example, the multivariate normal perturbation kernel using the full estimated covariance of the whole previous population has an acceptance rate similar to the component-wise normal perturbation kernel with adaptive estimation of the variances (see Figure \ref{fig:banana} B). The multivariate normal kernel with OLCM obtains slightly better results. The two versions of the perturbation kernel based on the FIM have significantly different acceptance rates. The most efficient version in term of acceptance rate is the one using the $M$ nearest neighbours, as might be expected. However this kernel, as in the case of the multivariate normal kernel based on the $M$ nearest neighbours, can show undesirable dependence on the chosen value of $M$. Figure~\ref{fig:banana} B represents the acceptance rate evolution for this last perturbation kernel with different values of $M$. The acceptance rate diminish considerably as  $M$ increases and the kernel becomes less ``locally aware''.

\section{Applications in Molecular Biology}
\subsection{The Repressilator Model}
\begin{figure}
\centering
 \includegraphics[width=\textwidth]{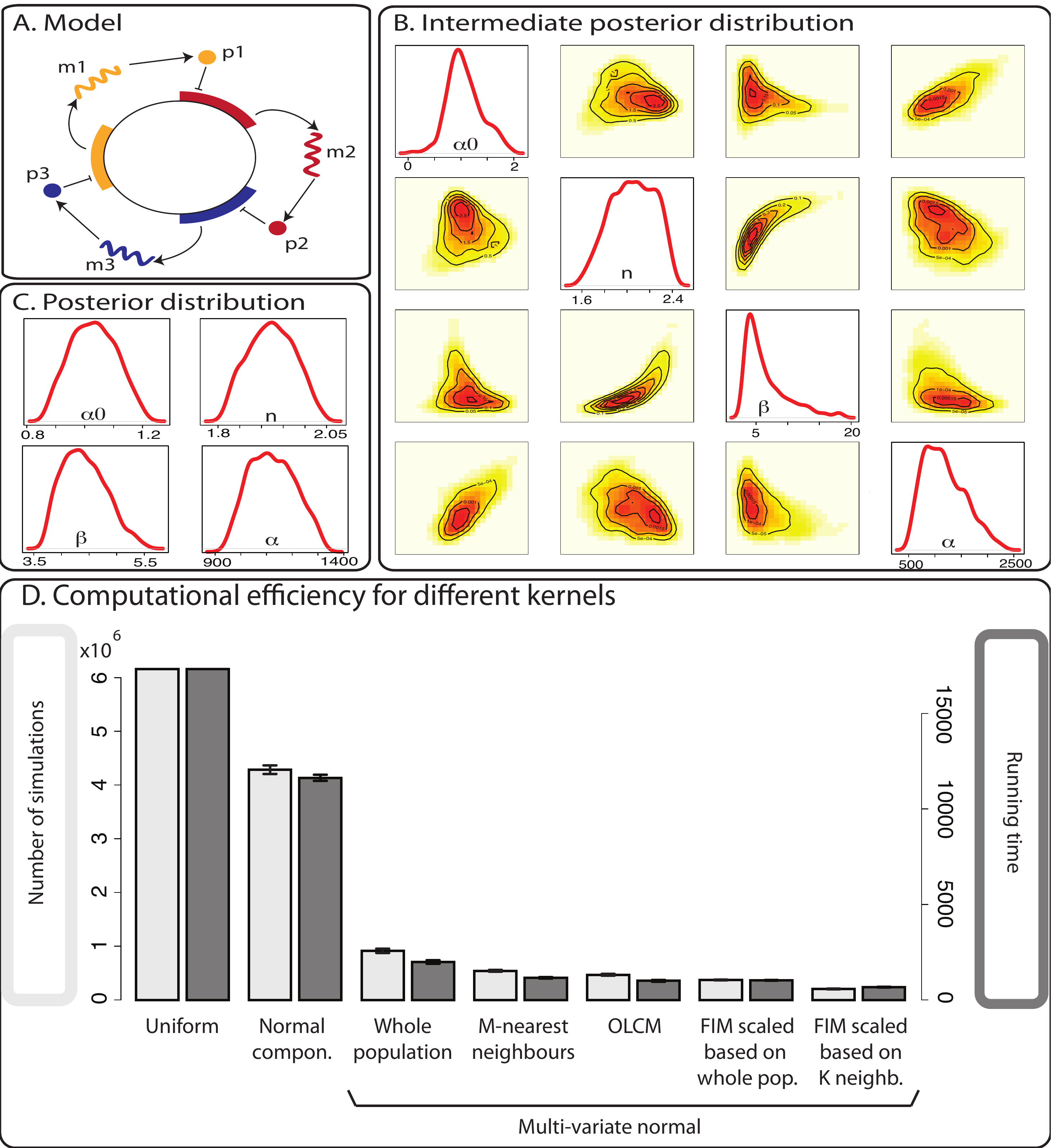}
\caption{A. The repressilator model: three genes are connected in a feedback loop; each gene $i$, transcribes the repressor protein $p_i$ for the next gene in the loop. B. Posterior distribution for $\epsilon=50$. C. Marginal posterior distribution for $\epsilon=35$. D. Number of simulations required to obtain the posterior distribution and running time of the algorithm (in minutes)  for each kernel. The error bars on the barplots show the variance of the number of simulation and running time over $10$ (resp. $3$) independent simulations for the multi-variate normal kernels (resp. for the normal componentwise kernel). The mRNA measurements are given in the supplementary take 1.}% $\{m_1(k)\}_k=( 0, 2.04, 32.29, 4.13, 2.15, 5.09, 1.07, 3.67, 39.01, 73.83, 8.54, 17.62, 11.96)$, $\{m_2(k)\}_k=(0, 28.99, 32.29, 10.61, 55.27, 9.49, 68.56, 10.62, -1.95, 3.53, 63.87, 39.68, -0.6)$ and $\{m_3(k)\}_k=( 0, 20.96, 7.49, 44.25, 7.12, 60.52, 8.10, 63.76, 22.9, 6.27, 10.59, 6.50, 70.56)$}
\label{fig:Repressilator}
\end{figure}
To analyse the differences between the efficiency of the perturbation kernel in biological applications we first focus on simulated datasets for the repressilator model, a popular model for gene regulatory systems  \citep{Elowitz:2000p13}. It consists of three genes connected in a feedback loop, where each gene transcribes the repressor protein for the next gene in the loop (see Figure~\ref{fig:Repressilator} A.). This model also exemplifies the challenges that are frequently encountered in attempts to reverse engineer the structure and parameters of dynamical systems from data \citep{Toni:2009p9197,Girolami:2009wl}. 
%There are now several studies that appear to demonstrate that, even for large data-sets, only about a third  of parameters of dynamical systems can be inferred with high confidence (or high posterior probability) \citep{Gutenkunst:2007,Rand:2008,Erguler:2011bu}; there are also signs, however, that judiciously chosen experimental conditions can lead to an increased information content in the data \citep{Casey:2007,Apgar:2008gh}.
\par
The evolution of the concentration of the $3$ proteins and mRNA over time is described by a system of six ordinary differential equations parametrized by a four dimensional parameter vector, $\theta=(\alpha_0,n,\beta,\alpha)$ as follows
\begin{align*}
&\frac{dm_1}{dt}=-m_1+\frac{\alpha}{1+p_3^n}+\alpha_0\\
&\frac{dp_1}{dt}=-\beta(p_1-m_1)\\
&\frac{dm_2}{dt}=-m_2+\frac{\alpha}{1+p_1^n}+\alpha_0\\
&\frac{dp_2}{dt}=-\beta(p_2-m_2)\\
&\frac{dm_3}{dt}=-m_3+\frac{\alpha}{1+p_2^n}+\alpha_0\\
&\frac{dp_3}{dt}=-\beta(p_3-m_3)\;.\\
\end{align*}
We denote by $m_i$ and $p_i$ the concentration of the mRNA and protein products of gene $i$ respectively. The parameters in the model are the Hill coefficient  $n$, repression strength $\alpha$, basal expression rate $\alpha_0$ and the ratio of the protein decay rate to the mRNA decay rate $\beta$. These are assumed to be the same for all three genes.
\par
We assume that only the mRNA $(m_1, m_2, m_3)$ measurements are available, and set the initial species concentrations
of $(m1, p1,m2, p2,m3, p3)$ to  $(0,2,0,1,0, 3)$; data are generated by simulating
the model with $(\alpha_0,n,\beta,\alpha)= (1,2,5,1000)$, and measuring the concentration of mRNA at time-points
$(0.0,0.6,4.2,6.2,8.6,13.4,16,21.4,27.6,34.4,39.8,40.6,45.2)$, subject to some added zero-mean Gaussian noise with variance 5. The same problems obviously prevail for different datasets.
\par
The ABC SMC algorithm is used to estimate $p(\theta|\{m_1(k), m_2(k),m_3(k)\}_k)$ with $N=1000$ particles and a decreasing sequence of thresholds equal to $(160,150,140,130,120,100,80,50,40,37,35)$. The marginal posterior distribution is represented in Figure \ref{fig:Repressilator} C and agrees very well with what is known from previous studies \citep{Toni:2009p9197}. An intermediate posterior distribution is represented in Figure \ref{fig:Repressilator} B and shows a highly non-linear correlation between some parameters, in particular parameters $n$ and $\beta$.
\par
In Figure~\ref{fig:Repressilator} D, we compare the cumulative number of sampled data over the algorithm as well as the running time of the algorithm
for different perturbation kernels. 
Using the uniform kernel, up to $6\times10^6$ simulations are required to obtain an approximation of the posterior distribution whereas less than $1\times10^6$ simulations are required if a multi-variate normal kernel is used with, in particular, only $1.6\times10^5$ simulations if the second version of the multivariate normal kernel based on the FIM is used. Moreover, we observe that the running time of the algorithm for each kernel is proportional to the number of simulations, so the main computational cost of the algorithm is due to the simulation of the data. Therefore, the time spent on defining the kernels, including determining the nearest neighbours of each particle, or evaluating the FIM for each parameter is small compared to the time saved by proposing new particles more efficiently.  So even in this simple model a significant improvement is possible through the appropriate choice of perturbation kernel.

\subsection{The Hes1 model}
\begin{figure}
\centering
 \includegraphics[width=\textwidth]{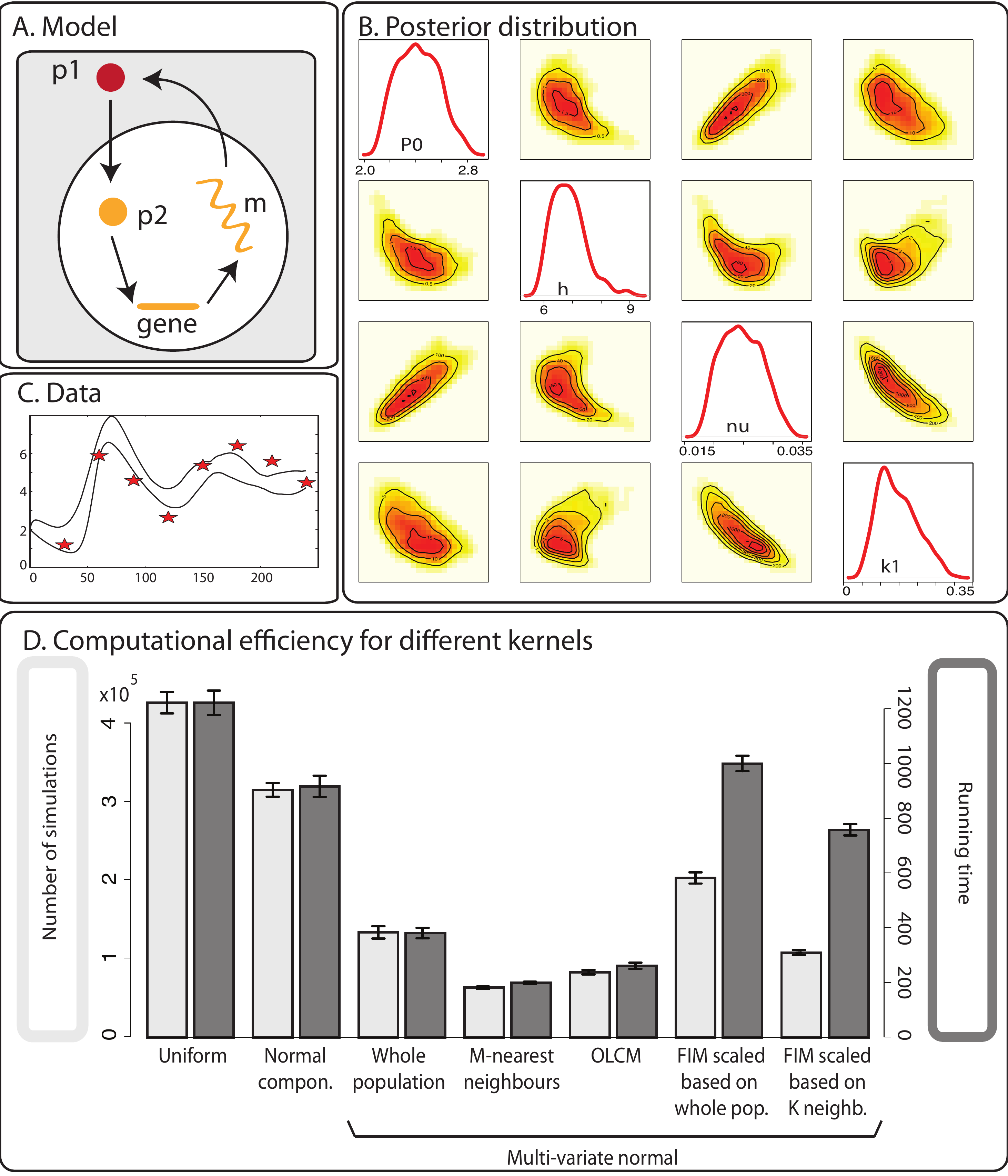}
\caption{A. The Hes1 model: the Hes1 protein migrates from the cytoplasm to the nucleus where it regulates the transcription of the mrna which then translates the cytoplasmic protein. B. Posterior distribution. C. The time-course of mRNA concentration: the stars represent the data published in \cite{silk2011designing}, the lines represent the minimum and maximum values for the evolution of the species for $1000$ particles sampled from the posterior distribution. D. Number of simulations required to obtain the posterior distribution and running time of the algorithm (in minutes)  for each kernel. The error bars on the barplots show the variance of the number of simulations and running time over $10$ independent simulations for each perturbation kernel. }
\label{fig:hes1}
\end{figure}

We now compare the efficiency of the perturbation kernels on an experimental dataset. We consider a dynamical system describing the expression level of the transcription factor Hes1, which plays an important role in cell differentiation and the segmentation of vertebrate embryos. In 2002, oscillations of Hes1 expression level as been observed by \cite{hirata2002oscillatory}. The Hes1 oscillator can be modelled by a system of three ordinary differential equations describing the evolution of the Hes1 mRNA, $m$, the Hes1 cytosolic protein, $p1$, and the Hes1 nuclear protein, $p2$ as follows (see Figure~\ref{fig:hes1} A and \cite{silk2011designing})
\begin{align*}
&\frac{dm}{dt}=-k_{deg}m+\frac{1}{1+(p_2/P_0)^h}\\
&\frac{dp_1}{dt}=-k_{deg}p_1+\nu m -k_1 p_1\\
&\frac{dp_2}{dt}=-k_{deg}p_2+k_1p_1\;.
\end{align*}
The model is parametrized by the protein degradation rate $k_{deg}$ which is assumed to be the same for both cytoplasmic and nuclear proteins, the rate of transport $k_1$ of Hes1 protein into the nucleus, the amount $P_0$ of Hes1 protein in the nucleus, when the rate of transcription of Hes1 mRNA is at half its maximal value,  the rate $\nu$ of  translation of Hes1 mRNA and the Hill coefficient $h$.
\par
To infer the parameters of this model we use the qualitative real-time PCR measurements published by \cite{silk2011designing}: the concentration of mRNA is measured every 30 minutes over 3 hours and is equal to $\{m\}_k=(2, 1.20, 5.90, 4.58, 2.64, 5.38, 6.42, 5.60, 4.48)$ (see Figure~\ref{fig:hes1}). We aim at inferring the $4$ parameters $\{P_0, \nu, k_1, h\}$; we take the degradation rate $k_{deg}$ is equal to be the experimentally determined value of $0.03$. According to the data and some preliminary results, we fix the initial concentration of mRNA to be equal to $2$, the initial concentration of $p_1$ to $5$ and the one of $p_2$ to $3$. The ABC SMC algorithm is used to estimate the posterior probability distribution with $N=1000$ particles and a decreasing sequence of thresholds equal to $(20, 13, 10, 6, 5, 4, 3, 2.8, 2.7, 2.6, 2.5)$. The posterior distribution is represented in Figure~\ref{fig:hes1} and is identical for every perturbation kernel.
To compare the performance of the perturbation kernels we show in Figure~\ref{fig:hes1} C the number of simulations during the algorithm as well as the running time of it for each kernel independently. We observe that the multi-variate normal kernels outperforms the uniform and normal componentwise kernels with an increase of the speed up to 4 folds for the kernel based on the M-nearest neihbours with $M=50$. 
Contrary to the repressilator model studied in the previous section, in the Hes1 dynamical system, the kernels based on the FIM do not perform well, with a running time comparable to the normal component-wise kernel. This may be explained by the fact that we use real data which may not be entirely explained by the model. In such a situation the likelihood surface may not be smooth enough for the FIM to offer much help. By contrast, however, the performance of the OLCM perturbation kernel is similar to that of the $50$-nearest neighbours kernel, with the advantage of being free of any parameter choice. For such a model we would clearly recommend the use of the OLCM perturbation  kernel.

\section{Conclusion}
In contrast to MCMC, where the pivotal role of perturbation kernels for convergence and mixing has been well documented \citep{Gilks1996,Robert2004}, for ABC SMC approaches there has been comparatively little work. In particular in the ABC context, which often relies on computationally costly simulation routines, poor choice of the perturbation kernel will result in potentially prohibitive computational overheads. We have addressed this lack of suitable kernels here in a rigorous but non-exhaustive fashion by focusing on kernels that are based around uniform or normal/multivariate normal parametric families. Importantly, in all the examples we were able to ensure that the different kernels had arrived at essentially identical posterior distributions, and for fixed $\epsilon_t$ schedule we can use the acceptance rate as an objective criterion for the numerical efficiency of different kernels.
\par
For all these models it is relatively straightforward to construct optimality criteria by reference to the KL divergence following \cite{Beaumont:2009be}. In higher-dimensional parameter spaces it is important to take into account the potentially correlated nature of parameters, and, not surprisingly we find that component-wise perturbation of particles tends to perform poorly compared to the other approaches considered here. In more complicated cases, e.g. decidedly non-Gaussian posteriors, multimodal posteriors, or posteriors with ridges, we find that a straightforward multivariate normal kernel is in turn inferior to kernels that are conditioned on the local environment of a particle. 
\par
In most applications of interest, the computational cost of simulating the data exceeds the algorithmic complexity $O(N^2)$ of the ABC SMC scheme. We therefore argue that the choice of a kernel with a high acceptance rate enables users to optimize the computational cost. However, when two kernels have the same acceptance rate --- which may happen for some shapes of the posterior --- it is more appropriate to select the one which is cheaper in terms of algorithmic complexity. The following table summarizes the computational cost of implementing the proposed perturbation kernels from a previous population of  $N$ particles with dimension $d$ (the number of individual parameters). In the case of the multivariate normal kernel based on the FIM, we denote by $C$ the computational cost of simulating an observation, \eg by solving the set of ODEs or SDEs which define the generative model. 
\par
\begin{tabular}{|c|c|}
\hline
Component-wise normal& $O(dN^2)$\\
Multivariate normal based on the whole previous population & $O(d^2N^2)$\\
Multivariate normal based on the $M$ nearest neighbours & $O((d+M)N^2+d^2M^2N)$\\
Multivariate normal with OLCM & $O(d^2N^2)$\\
Multivariate normal based on the FIM & $O(dCN+d^2N^2)$\\
 (normalized with entire population)&\\
\hline
\end{tabular}
\\
\par
As a general rule of thumb we would recommend the use of multi-variate kernels with OLCM, which tends to have the highest acceptance rate in our examples and is relatively easy to implement at acceptable computational cost. 
For some probability models, in particular those describing dynamical systems, the FIM has attracted a lot of attention recently \citep{Amari:2007wt, Arwini:2008ur,Secrier:2009ko,Girolami:2009wl,Erguler:2011bu,Komorowski:2011cn}, and it appears likely that we will be able to exploit these notions, and those of information geometry more generally, fruitfully in ABC SMC. Therefore, where applicable, we tested the use of the FIM to drive the perturbation kernel. Theoretically, and as can be seen in the Repressilator model, this has the advantage of exploring so-called neutral spaces more efficiently while maintaining a high acceptance rate. However, the results on the Hes1 model suggest that the perturbation kernels based on the FIM are not robust enough and can lead to high computational cost in particular for relatively complex biological models with real data which induce a non-smooth likelihood surface.
%However there is a potential tension between global and local aspects; if we take the global FIM as the basis for the perturbation we may have a very poor representation of the internal structure of \eg the posterior (very much as might happen in the Laplace expansion); if, on the other hand, we evaluate the FIM based on the $M$ nearest neighbours of a particle, then we may overly restrict the particles making up the next intermediate distribution. 
\par
The cost of local measures based on $M$ nearest neighbours may seem too high to contemplate their use. However, given the increase in acceptance rate that we have observed, and the generally high computational cost of simulating complex data, they can prove to be fruitful. The user should be aware that the efficiency of such a measure strongly depends on the chosen value of $M$ (the smaller the value of $M$ the higher the acceptance rate) and that a too small value of $M$ can lead to a too local perturbation kernel with a risk of not converging to the posterior distribution as $\epsilon$ goes to $0$. This sensitivity to a user-defined parameter makes the $M$ nearest neighbours kernel less user-friendly than OLCM.
\par
The kernels discussed here may seem restrictive, especially to those from a background in evolutionary computation. We may, for example, wish to consider other perturbations to generate new candidate particles, such as recombination \citep{Baragona:2010aa}, as is frequently done in global optimization. In principle it is possible to include this in ABC SMC approaches, as long as the weights for new particles can be calculated (which turns out to be relatively straightforward for recombination and different cross-over schemes). It has to be kept in mind, however, that these perturbations work best in cases where the parameter space is so under-sampled that random combinations of individual parameters are sufficiently likely to end up in a region with a more favourable cost-function than a local, \eg gradient-based proposal would. While such strategies have been applied in many optimization settings, their use in Bayesian inference is rare, since generally here the optimum (by whichever criterion) parameter value is of less interest than the distribution as a whole. For maximum a posteriori inferences such methods may be fruitfully applied, but here we do not see an obvious advantage (as is also borne out by simulation studies, data not shown). 
\par
Kernel choice is one of the obvious means of speeding up ABC SMC inferences. Setting the $\epsilon$ schedule optimally is another. The latter is straightforwardly automated by basing the next $\epsilon_{t+1}$ on the acceptance rate obtained during the generation of the intermediate distribution, $p_{\epsilon_t}(\theta|x)$. But again there is a trade-off to be made between convergence and exhaustive exploration of the parameter space. In particular too gentle a decrease in $\epsilon_t$ may result in loss of particle diversity \citep{Silk:Filippi:2012}. Here we believe that further investigation of FIMs may hold important clues to how the $\epsilon_t$ are best chosen. This would, for example, resonate with the perspective on $\epsilon$ proposed by \cite{Ratmann:2009by}.

\subsection*{Acknowledgements} SF is funded through an MRC Computational Biology Research Fellowship; CB and MPHS gratefully acknowledge financial support from the BBSRC (BB/G007934/1); MPHS is a Royal Society Wolfson Research Merit award holder.

\bibliographystyle{interface}
%\bibliography{bibkernels.bib,/Users/michael/bibliography/mstbibnet.bib,/Users/michael/bibliography/wholebib.bib} 
\bibliography{biblionew}

\end{document}